\documentclass{egpubl}
\usepackage{eurovis2024}

\CGFccby

\usepackage[T1]{fontenc}
\usepackage{dfadobe}  

\usepackage{cite}  
\BibtexOrBiblatex
\electronicVersion
\PrintedOrElectronic
\ifpdf \usepackage[pdftex]{graphicx} \pdfcompresslevel=9
\else \usepackage[dvips]{graphicx} \fi

\usepackage{egweblnk}

\title[Operation and Manifestation in Dynamic Physicalizations]%
      {Investigating the Effect of Operation Mode and Manifestation\\on Physicalizations of Dynamic Processes}

\author[D. Pahr \& H. Ehlers \& H. Wu \& M. Waldner \& R. G. Raidou]
{\parbox{\textwidth}{\centering Daniel Pahr$^1$\orcid{0000-0001-7313-3056}, Henry Ehlers$^1$\orcid{0000-0002-5994-1492}, Hsiang-Yun Wu$^{2,1}$\orcid{0000-0003-1028-0010}, Manuela Waldner$^1$\orcid{0000-0003-1387-5132} and Renata\,G. Raidou$^1$\orcid{0000-0003-2468-0664} 
        }
        \\
{\parbox{\textwidth}{\centering $^1$TU Wien, Austria\\
         $^2$St. P{\"o}lten University of Applied Sciences, Austria
       }
}
}

\usepackage{mathptmx}                  
\usepackage{caption}
\usepackage[table]{xcolor}
\usepackage{subcaption}
\usepackage{multirow}
\usepackage{graphicx}
\usepackage{array}
\usepackage{todonotes}
\usepackage{soul}
\usepackage[margin=1mm]{subcaption}
\usepackage[normalem]{ulem}
\usepackage[subtle]{savetrees}
\usepackage{lineno}

\usepackage{hyperref}
\usepackage{xurl}

\definecolor{physical}{HTML}{00AEEF}
\definecolor{virtual}{HTML}{F7941D}
\definecolor{manual}{HTML}{2BB673}
\definecolor{automatic}{HTML}{EC008C}

\newcommand{\add}[1]{{\textcolor{blue}{#1}}}

\newcommand{\pc}[1]{\setulcolor{physical}\ul{#1}}
\newcommand{\vc}[1]{\setulcolor{virtual}\ul{#1}}
\newcommand{\ac}[1]{\setulcolor{automatic}\ul{#1}}
\newcommand{\mc}[1]{\setulcolor{manual}\ul{#1}}
\newcommand{\pcm}[1]{{#1}}
\newcommand{\vcm}[1]{{#1}}

\newcommand{\mcm}[1]{{#1}}

\newcommand{\PM}[0]{\textbf{\setulcolor{physical}\ul{\texttt{P}}\setulcolor{manual}\ul{\texttt{M}}}}
\newcommand{\PA}[0]{\textbf{\setulcolor{physical}\ul{\texttt{P}}\setulcolor{automatic}\ul{\texttt{A}}}}
\newcommand{\VM}[0]{\textbf{\setulcolor{virtual}\ul{\texttt{V}}\setulcolor{manual}\ul{\texttt{M}}}}
\newcommand{\VA}[0]{\textbf{\setulcolor{virtual}\ul{\texttt{V}}\setulcolor{automatic}\ul{\texttt{A}}}}
\newcommand{\Ph}[0]{\textbf{\setulcolor{physical}\ul{\texttt{P}}}}
\newcommand{\Vi}[0]{\textbf{\setulcolor{virtual}\ul{\texttt{V}}}}
\newcommand{\Ma}[0]{\textbf{\setulcolor{manual}\ul{\texttt{M}}}}
\newcommand{\Au}[0]{\textbf{\setulcolor{automatic}\ul{\texttt{A}}}}

\usepackage{caption}
\begin{document}

\teaser{
  \centering
  \includegraphics[width=0.8\linewidth]{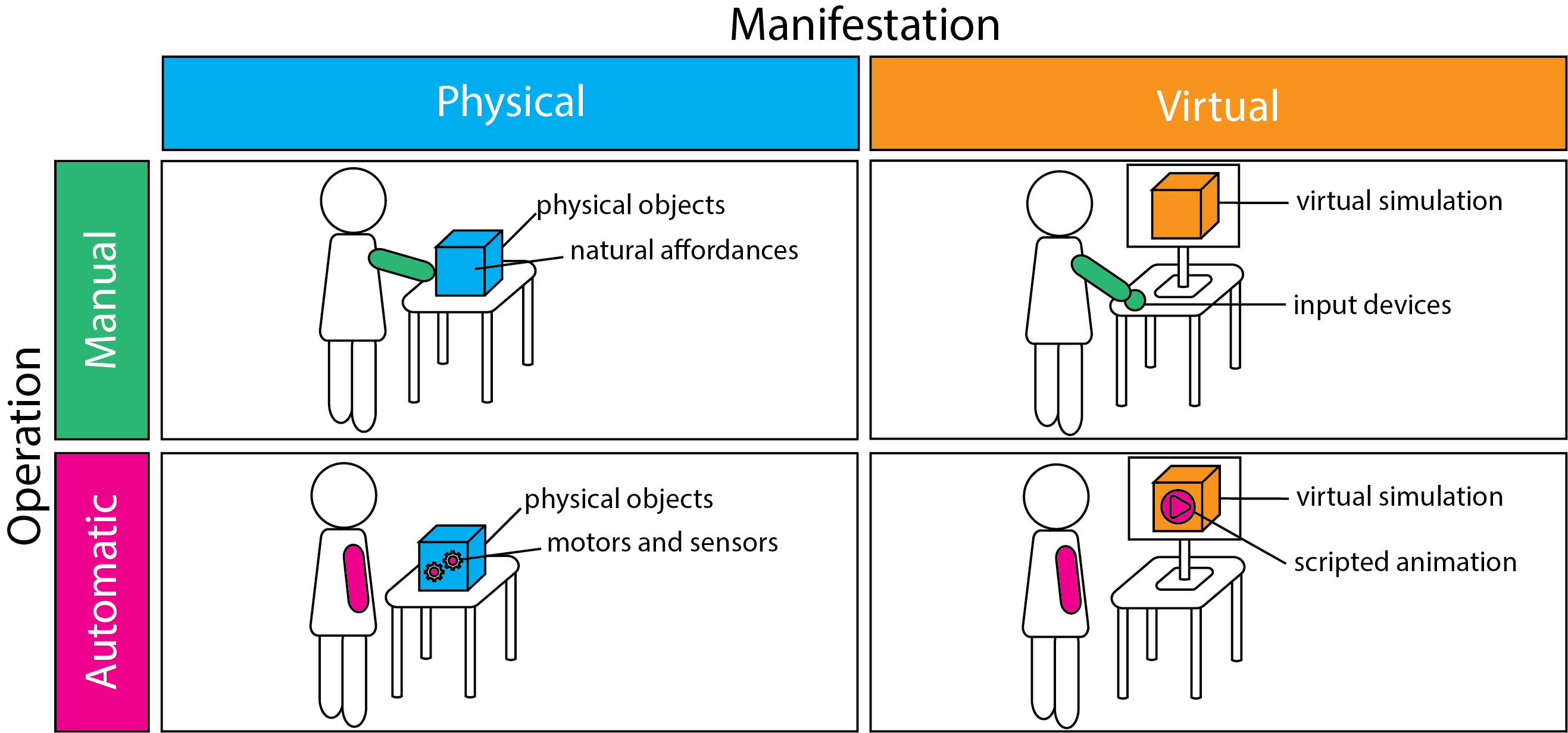}
  \caption{The two-dimensional design space explored in our study delves into the communication of complex dynamic processes. The two dimensions comprise the \emph{manifestation} of a representation (\setulcolor{physical}\ul{physical} or \setulcolor{virtual}\ul{virtual}) and the mode of \emph{operation} (\setulcolor{manual}\ul{manual} or \setulcolor{automatic}\ul{automatic}).}
  \label{fig:teaser}
}

\maketitle
\begin{abstract}
We conducted a study to systematically investigate the communication of complex dynamic processes along a two-dimensional design space, where the axes represent a representation's manifestation (\setulcolor{physical}\ul{physical} or \setulcolor{virtual}\ul{virtual}) and operation (\setulcolor{manual}\ul{manual} or \setulcolor{automatic}\ul{automatic}).
We exemplify the design space on a model embodying cardiovascular pathologies, represented by a mechanism where a liquid is pumped into a draining vessel, with complications illustrated through modifications to the model.
The results of a mixed-methods lab study with 28 participants show that both physical manifestation and manual operation have a strong positive impact on the audience's engagement. 
The study does not show a measurable knowledge increase with respect to cardiovascular pathologies using manually operated physical representations.
However, subjectively, participants report a better understanding of the process---mainly through non-visual cues like haptics, but also auditory cues.
The study also indicates an increased task load when interacting with the process, which, however, seems to play a minor role for the participants. 
Overall, the study shows a clear potential of physicalization for the communication of complex dynamic processes, which only fully unfold if observers have to chance to interact with the process.


\end{abstract}

\section{Introduction}
\label{sec:intro}

Visual representations, whether in virtual or physical form, are integral tools in education, widely embraced for their effectiveness. 
In STEM fields, the use of visual representations has been extensively studied with concepts like pulleys in physics \cite{gire_effects_2010, ferguson_learning_1995, boucheix_static_2009}, prompting an exploration into the factors contributing to their widespread adoption. 
Among others, interaction with these representations is one of the most documented elements as it enhances the audience's understanding.

In the medical context, conveying (patho)physiological processes poses unique challenges compared to more straightforward concepts in physics, such as pulleys. 
For instance, communicating the process behind a ``healthy'' cardiac cycle to \add{laypeople} involves explaining the coordinated phases of systole and diastole, the role of heart valves and chambers, and the efficient circulation of blood~\cite{waugh_ross_2022}.
Communicating potential pathophysiologies, i.e., functional changes accompanying a pathological condition, adds complexity as it involves comprehending the impact of disruptions on heart function and overall health.
Yet, patient communication and education are crucial for informed decision-making, active participation in healthcare, and an enhanced overall patient understanding of their conditions and treatment options~\cite{meuschke_towards_2021}.

Within the realm of visual representations, physicalizations emerge as a unique opportunity for bringing data into the \pcm{physical} space~\cite{dragicevic_data_2021}. 
This is primarily due to the distinct characteristics of physical representations, such as the use of physical embodiment and natural affordances to convey meaning and engage an audience \cite{zhao_embodiment_2008}.
While physicalizations find widespread use in STEM~\cite{gire_effects_2010}, their application in explaining dynamic processes of higher complexity remains relatively unexplored \cite{rau_comparing_2020}.

Recent work in data physicalization targeted the creation of engaging and low-cost anatomical models for edutainment and patient education---without addressing pathophysiological or any other dynamical aspects~\cite{stoppel_vol2velle_2017, pahr_vologram_2021, schindler_nested_2022}. 
All prior examples employ \textit{indirect interaction} that acts similar to interface controls \cite{bae_making_2022}.
Conversely, multisensory displays \cite{hogan_towards_2016} use \textit{direct interaction}, directly stimulating a user's sensory affordances \cite{bae_making_2022}. 
The concept of direct interaction opens up additional channels---for example, the use of kinaesthetics---to encode data into physical activity performed by observers~\cite{hurtienne_move_2020}. 
Directly \mcm{interactive} models hold promise as representations of pathophysiological processes, but they remain largely unexplored and unassessed in this context.

Research in data physicalization has sought to quantify the value of \pcm{physical} data representations~\cite{jansen_evaluating_2013, stusak_evaluating_2015}. 
This has often been done using comparative methods, where \pcm{physical} data representations are compared against similar screen-based ones. 
Until recently, evaluations of both \pcm{physical} and \vcm{virtual} data representations focused primarily on efficiency, measuring how quickly insights can be derived.
This narrow focus has inadvertently left unexplored---or even obscured---potential advantages of data physicalization, especially for the medical domain where audience engagement in the communication of complex processes is crucial\cite{meuschke_towards_2021}.
Recently, new metrics with a focus different from efficiency or comprehension have emerged offering to support the evaluation of such concepts~\cite{wang_emotional_2019}.

To fill this gap in research, we investigate the potential of directly interactive models that represent dynamic pathophysiological processes. We also delve into assessing the implications and benefits of data physicalization beyond traditional efficiency metrics.
To this end, we explore the effects of \textbf{manifestation} and mode of \textbf{operation} of a representation on the communication of pathophysiological processes on \textbf{observer understanding (Q1)}, \textbf{subjective task load (Q2)}, and \textbf{enjoyment (Q3)}.
We consider the manifestation---\vc{virtual} or \pc{physical}---and the mode of operation---\mc{manual} or \ac{automatic}---as two individual factors in a full factorial study with 28 participants. Figure~\ref{fig:teaser} shows an overview of the resulting design space. We employ a \textbf{mixed methods} study design. First, we use quantitative methods to validate a-priori-postulated hypotheses about understanding, task load, and emotional engagement. Subsequently, we analyze the qualitative feedback of our study subjects. Finally, using triangulation \cite{ocathain_three_2010} we gather the insights obtained through both methods, to draw further conclusions on our findings. 

\noindent\textbf{The contribution} of our work stems from the results of an exploration of a two-dimensional design space for educational models for pathophysiological processes (Figure~\ref{fig:teaser}).
We confirm that both \pc{physical} manifestation and \mc{manual} operation increase engagement.
Qualitative feedback reveals that \mc{manual} interaction with \pc{physical} representations augments non-visual stimuli, leading to subjective knowledge increase.
With the results of this study, we discuss the implications that arise from our design space in educational pathophysiological process representations.
\section{Related Work}
\label{sec:related}

\noindent{\textbf{Representations for Learning}} Numerous studies in educational sciences have explored the use of physical and virtual representations, examining their impact on learning.
Ferguson and Hegarty~\cite{ferguson_learning_1995}compare learning effects achieved with physical representations of a mechanical system to diagrams abstracting the underlying concepts. They found that participants who were provided physical models were able to solve application tasks more accurately. 
Similarly, Boucheix and Schneider~\cite{boucheix_static_2009} compared static and animated learning aides. They showed a positive influence of animations compared to static data representations for dynamic systems. Yet, an influence of user control on comprehension was not indicated.
Finally, Gire et al.~\cite{gire_effects_2010} investigated the difference between physical and virtual learning aides. A virtual simulation and a physical model were employed in a comparative study to determine whether either type of manifestation provided advantages. They found certain aspects of learning were improved by physical interaction.

Han~\cite{han_embodiment_2013} used the example of gearbox transmission to evaluate the influence of physicality on learning. Using a Lego model and a computer simulation, they investigated if either form of manifestation would help understand the concept better. While neither model provided a significant advantage, they found that participants who had used a car with a manual gearbox performed better in the experiment. This could indicate a benefit of haptic feedback on the learning process.
In a recent literature review, Rau~\cite{rau_comparing_2020} highlights gaps in the comparison of learning with virtual and physical representations. They indicated potential in researching the benefit of added haptic feedback to a physical representation and suggested isolating mechanisms in models to investigate this. Furthermore, they recommended accounting for cognitive load effects in the research of physical representations, and the investigation of the impact of physical engagement on learning outcomes.

Both physical and virtual representations have been employed in the context of learning, and all aforementioned examples use a direct representation of a mechanical concept. 
In contrast, our focus extends to dynamic pathophysiological systems, offering a novel perspective on complex structures and functional interplays within learning contexts.

\noindent{\textbf{Physicalization and Medical Education}} In recent years, research in the fields of digital fabrication and data physicalization has presented novel ways to transform traditional medical education.
Ang et al.~\cite{ang_physicalizing_2019} describe a 3D-printed representation of the blood flow in the human heart. 
Utilizing flow lines and glyphs, they allow observers to interactively manipulate the model, providing a detailed exploration of specific anatomical regions.
While the ever-growing field of data physicalization has produced many examples of creative data representations~\cite{djavaherpour_data_2021}, only a few works focus on its application to the education of the general public or lay audiences.
Stoppel and Bruckner~\cite{stoppel_vol2velle_2017} propose a method that uses transparent rotating discs to create custom illustrations of medical volume data and playfully explore---otherwise complex to explain---rendering settings.
Similarly, Raidou et al.~\cite{raidou_slice_2020} and Pahr et al.~\cite{pahr_vologram_2021} present different ways to create anatomical sculptures from transparent materials, which when assembled in 3D support anatomical education. Using foldable papercrafts, Schindler et al.\cite{schindler_anatomical_2020, schindler_nested_2022} describe a workflow for the creation of nested and multilayered paper models for anatomy education. 

Norooz et al. \cite{norooz_bodyvis_2015} design a wearable approach to anatomy education aimed at children. Layered, detachable organs show where and how organs are located within the body. Study sessions together with educators highlighted the potential of this method to provoke curiosity.
All aforementioned examples are designed around displaying medical imaging data and are limited to indirect interactivity.

\noindent{\textbf{Direct Interaction with Data Physicalization}}
Bae et al.~\cite{bae_making_2022} define two different types of interactions with data physicalizations: indirect and direct. Indirect interactions control a physicalization through a user interface, while direct interactions directly stimulate the user's senses through its affordances. Bae et al. \cite{bae_computational_2024} present an approach to produce interactive network visualizations. A central principle to their approach is that the physicalization also doubles as a tangible interface that allows node selection on a digital twin of the network. 
INTUIT \cite{djavaherpour_first_2023} shows a method to encode the roughness of arctic ice into the tactile properties of a physical model.
Perovic et al.~\cite{perovich_tactile_2023} analyze such haptic interactions with data physicalizations, by using fluorescent markers. After tactile interactions, UV photography of the target physicalization reveals where the observer touched the sculpture.
Bringing this concept to a larger scale, Karyda et al.~\cite{karyda_narrative_2021} investigate the impact of directly interactive physicalizations by making people engage with table-sized personalized installations, tailored to people's personalities. They show examples of a treadmill, a foosball table, and a one-string instrument.
With ``Move\&Find``, Hurtienne et al.~\cite{hurtienne_move_2020} embody the energy consumption of a Google search to a treadmill metaphor.

These examples present the integration of the observer as an active part of physical data representations. However, all aforementioned representations are designed to illustrate static datasets or facts and do not touch upon dynamic processes. 

\noindent{\textbf{Evaluating Physicality and Interactivity}}
Evaluation is an inherent component of all aforementioned works. In the field of data physicalization and multisensory data displays, common comparisons involve virtual and physical representations  \cite{jansen_evaluating_2013, stusak_evaluating_2015, pollalis_evaluating_2018, ang_physicalizing_2019}. 
Jansen et al.\cite{jansen_evaluating_2013} compare a physical 3D bar chart, virtual 2D, and 3D representations, finding that while the physical artifact outperforms the 3D virtual version, the 2D visualization is most efficient. Stusak et al.\cite{stusak_evaluating_2015} assess the memorability of physical vs. virtual bar charts, revealing better immediate memorization for the virtual chart but prolonged recall with the physical one.
Ang et al.\cite{ang_physicalizing_2019} compare a physicalized blood flow representation to a screen-based visualization, noting participants' faster interaction with the physical version. However, they find no clear evidence of superiority. The evaluation of "Move\&Find"\cite{hurtienne_move_2020} focuses on understanding, engagement, and behavioral change, revealing that the multisensory representation enhances data understanding, creativity, and engagement.
Drogemuller et al. \cite{drogemuller_haptic_2021} study the understanding of physical network representations. They compare physical representations both with and without tactile interaction, as well as virtual and haptic-only conditions. Their findings indicate heightened engagement, as well as self-perceived efficiency increase when people interact haptically with graph physicalizations. 
Pollalis et al. \cite{pollalis_evaluating_2018} evaluate the usefulness of 3D printed artifact replicas compared with virtual ones on a screen and an AR device in an archaeology education setting. They use a mixed methods approach, measuring task times, enjoyment, perceived task workload, spatial presence, and learning outcomes both qualitatively and quantitatively.   
Their study highlights shortcomings of 3D printed artifacts due to inaccurate reproduction during the printing process. 
Taher et al.\cite{taher_investigating_2017} and Sturdee et al. \cite{sturdee_exploring_2023} study interactions with an interactive physical bar chart in single \cite{taher_investigating_2017} and co-located user \cite{sturdee_exploring_2023} scenarios respectively. 
Their findings show the great potential of interactive data physicalizations to encourage exploration. The co-located study also indicates positive social engagement when groups interact with a physical display together. 
Sereno et al. \cite{sereno_hybrid_2022} compare different tangible interaction methods for the selection of data points in 3D space. Their findings indicate that AR devices have a lesser impact on subjective workload compared to screen-based methods. However, while users reported the interactions in AR to be more direct, this did not influence their performance.

Our study is the first to consider the manifestation of a representation and the way an observer interacts with it as independent factors. In doing so we avoid the confounding of beneficial effects of either factor. Furthermore, we individually assess how these factors influence not only learning but also task load and engagement.

\section{Methodological Approach}
\label{sec:study}

\noindent\textbf{Design Space} 
In this work, we investigate the effects of the \textbf{operation} mode and \textbf{manifestation} as separate factors within physicalizations of dynamic processes, such as those present in pathophysiology.
To this end, our design space is built upon two dimensions: the \textit{mode of operation}, describing the way the representation is operated, and \textit{manifestation}, describing the medium in which the representation exists. 
An observer operates the representation in a \mc{manual} way or the representation is \ac{automated}. 
The manifestation of the representation can be either \pc{physical} or \vc{virtual}. Our two-dimensional design space for process physicalizations is schematically represented in Figure~\ref{fig:teaser}.

Representation-based learning, as well as data physicalization, has previously investigated \textbf{physical manifestation} as a central concept. 
In the \pc{physical} \mc{manual} quadrant of our design space we consider only data physicalizations with \textbf{direct interactivity} \cite{bae_making_2022}. Examples, among the related works discussed in Section~\ref{sec:related}, are INTUIT \cite{djavaherpour_first_2023} or ``Move\&Find`` \cite{hurtienne_move_2020}. In the \vc{virtual} \mc{manual} quadrant, direct interaction is relegated to the input and output devices of, e.g., a PC. We choose not to add additional indirect interaction modalities to our virtual representations. Examples of indirect interactions from the related works of Section~\ref{sec:related} are Boucheix and Schneider's~\cite{boucheix_static_2009} controlled animations, or the visualizations used by Jansen et al.\cite{jansen_evaluating_2013} in comparison to their physical bar chart.
In the \ac{automated} quadrants, we remove the interaction with the representations completely. An example of a \vc{virtual} \ac{automatic} representation is Hurtienne et al.'s\cite{hurtienne_move_2020} animated cyclist, or other animated representations \cite{boucheix_static_2009}. Oppositely, Perovich et al.'s~\cite{perovich_chemicals_2021} floating lanterns representing environmental pollution are a \pc{physical} \ac{automatic} physicalization.

\noindent\textbf{Research Questions and Hypotheses} While prior work has not shown that physical representations are more efficient for information retrieval\cite{jansen_evaluating_2013, stusak_evaluating_2015, hurtienne_move_2020}, this has not been yet evaluated for representations of dynamic processes. Increased physical load interacting with physical representations has been observed in the past\cite{hurtienne_move_2020}. Finally, physical representations have been shown to be more engaging \cite{hurtienne_move_2020} than virtual ones. We, therefore, pose three separate research questions regarding \textbf{understanding (Q1)}, \textbf{task load (Q2)}, and \textbf{enjoyment (Q3)} of pathophysiological process representations---specifically, w.r.t. how they are impacted by different modes of operation and manifestation:

\noindent\textbf{H1:} Physical and manually operated representations are more effective in conveying processes than virtual and automated ones.

\noindent\textbf{H2:} Physical and manually operated representations entail a higher subjective task load than virtual and automated ones.    

\noindent\textbf{H3:} Physical and manually operated representations are more enjoyable than virtual and automated ones.

Metaphors~\cite{zhao_embodiment_2008} and visual abstractions~~\cite{viola_visual_2020} in the context of data communication serve to simplify intricate physiological processes, helping both professionals and the general public grasp complex concepts. 
To enhance comprehension for a lay audience, we introduce an abstract metaphor that elucidates a complex process within the human body: cardiac function. We discuss our metaphor choice and the respective representation design arising from it in Sections 3.1--2. 
Deriving from the design space, we create four representations, one for each quadrant.
These are physical manual (\PM), physical automated (\PA), virtual manual (\VM), and virtual automated (\VA).

Subsequently, to investigate our previously formulated hypotheses with regard to the conceptualized metaphor and its corresponding representation design, we employ both quantitative and qualitative methods. 
We initially look at how the mode of operation and manifestation influence an observer's understanding of a process representation (\textbf{H1}). We also measure individual task load (\textbf{H2}) and enjoyment, in terms of emotional engagement (\textbf{H3}), in interacting with said representations. Subjective preference will be used as an additional measure for enjoyment (\textbf{H3}). Written statements collected during the study will serve as the basis for our qualitative evaluation.

\subsection{Modeling Basic Cardiac Function with a Metaphor}

A primary goal for (physical) data representation is the communication of insights to laypeople with limited domain knowledge.
In the context of data visualization and data physicalization alike, both metaphors and visual abstractions serve as essential tools for conveying complex information to diverse audiences.
Metaphors in data physicalization function as powerful cognitive tools, enabling observers to relate to the representation of intricate processes and complex data sets by leveraging familiar concepts \cite{zhao_embodiment_2008}. 
Visual abstractions, on the other side, filter out unnecessary details in data representations allowing users to focus on crucial elements~\cite{viola_visual_2020}. 

We introduce an abstract metaphor to elucidate the complexities of the physiological processes of cardiac function. 
Cardiovascular diseases are the leading cause of global mortality, while many of the underlying risks can be addressed through behavioral changes\cite{roth_global_2020}.
Campaigns \cite{schulberg_cardiovascular_2022} and installations \cite{mcashan_health_2016} communicating cardiovascular function and associated risks are widely employed to raise awareness in the general population. Still, data physicalization approaches looking in this direction are scarce \cite{ang_physicalizing_2019} but are anticipated to facilitate a tangible and accessible comprehension of cardiovascular processes by transforming complex data into interactive and visually intuitive representations.
To ensure broad applicability, our model is designed for both virtual and physical representation, accommodating automated and manual operation. This versatility enhances the accessibility of medical information for laypeople, aligning with our goal of simplifying complex concepts in health education.

\noindent\textbf{The Pump Metaphor} In the circulatory system, the function of the heart is to pump blood through the circulatory system. During the diastole, the cardiac muscle relaxes and the chambers fill with blood. In the systolic phase, the muscle contracts, expelling the blood to the lungs and the peripheral vessels.
Besides the actual meaning that the heart transports fluids, this also provides us with an intuitive metaphor for its function.
Pumps are a familiar concept, inherently understandable to people.
Beyond the aforementioned metaphorical value of the subject, Offenhuber's perspective \cite{offenhuber_what_2020} gives additional directions.
The pump metaphor is very close to the actual function of the heart, defining the function in an \textit{ontological} sense. Laypeople may also not be familiar with the specific variables that medical professionals use when they speak about bodily functions, like blood oxygenation, blood pressure, or heart-stroke volume. We argue that a \textit{relational} perspective is better suited for non-experts.  
Deconstructing the heart function with the pump metaphor allows for a deeper understanding of the regulation process. Using an ontological-relational perspective limits the use of numeric representations with which a lay audience might not be familiar.

\begin{figure}[t]
    \begin{subfigure}{0.45\textwidth}
    \centering
    \includegraphics[width=\textwidth]{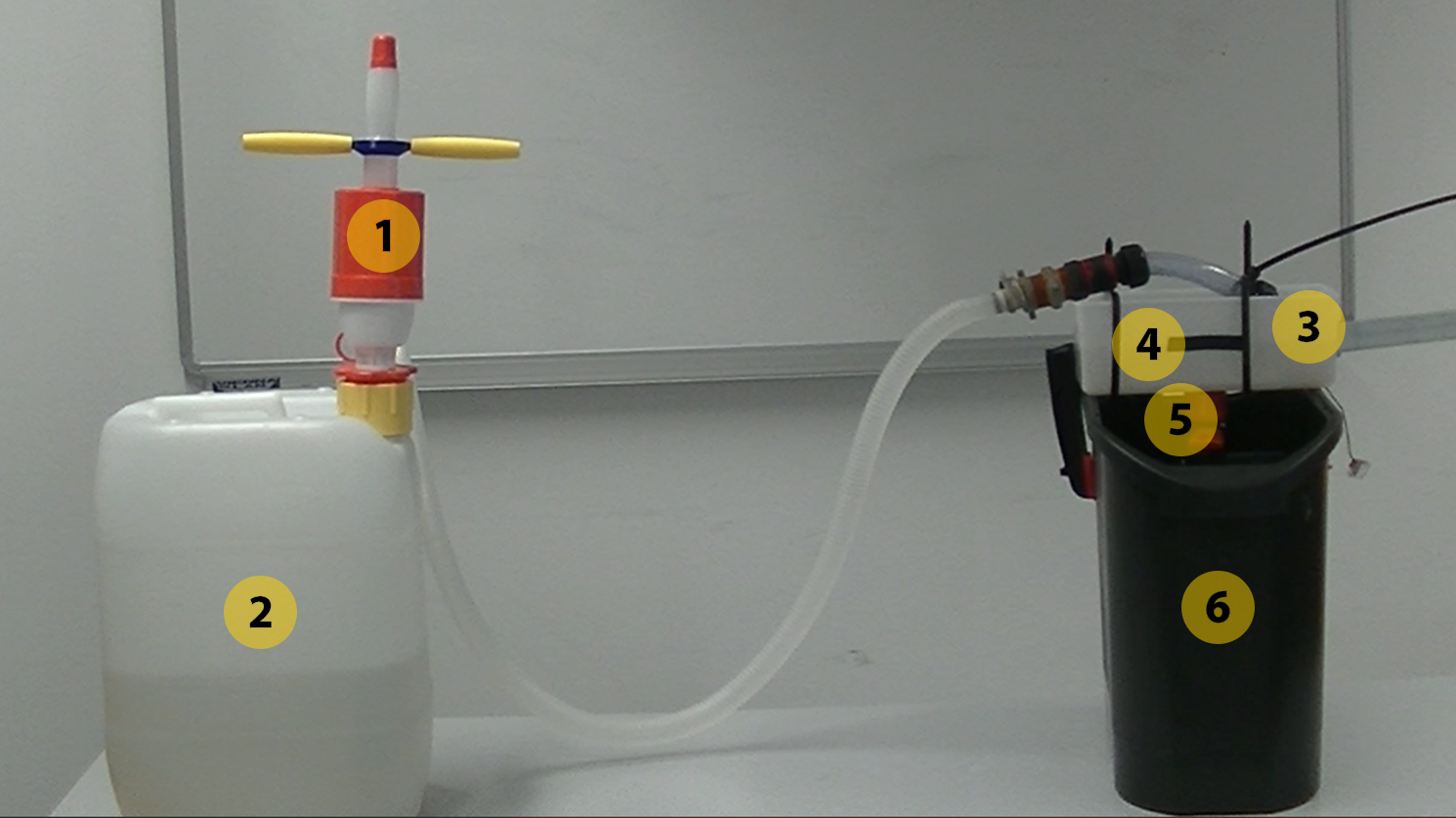}
    \subcaption{Physical. Our representation is embodied by physical objects, according to the metaphor.}
    \label{fig:setup_phys}
    \end{subfigure}
    \centering
    \begin{subfigure}{0.45\textwidth}
    \centering
    \includegraphics[width=\textwidth]{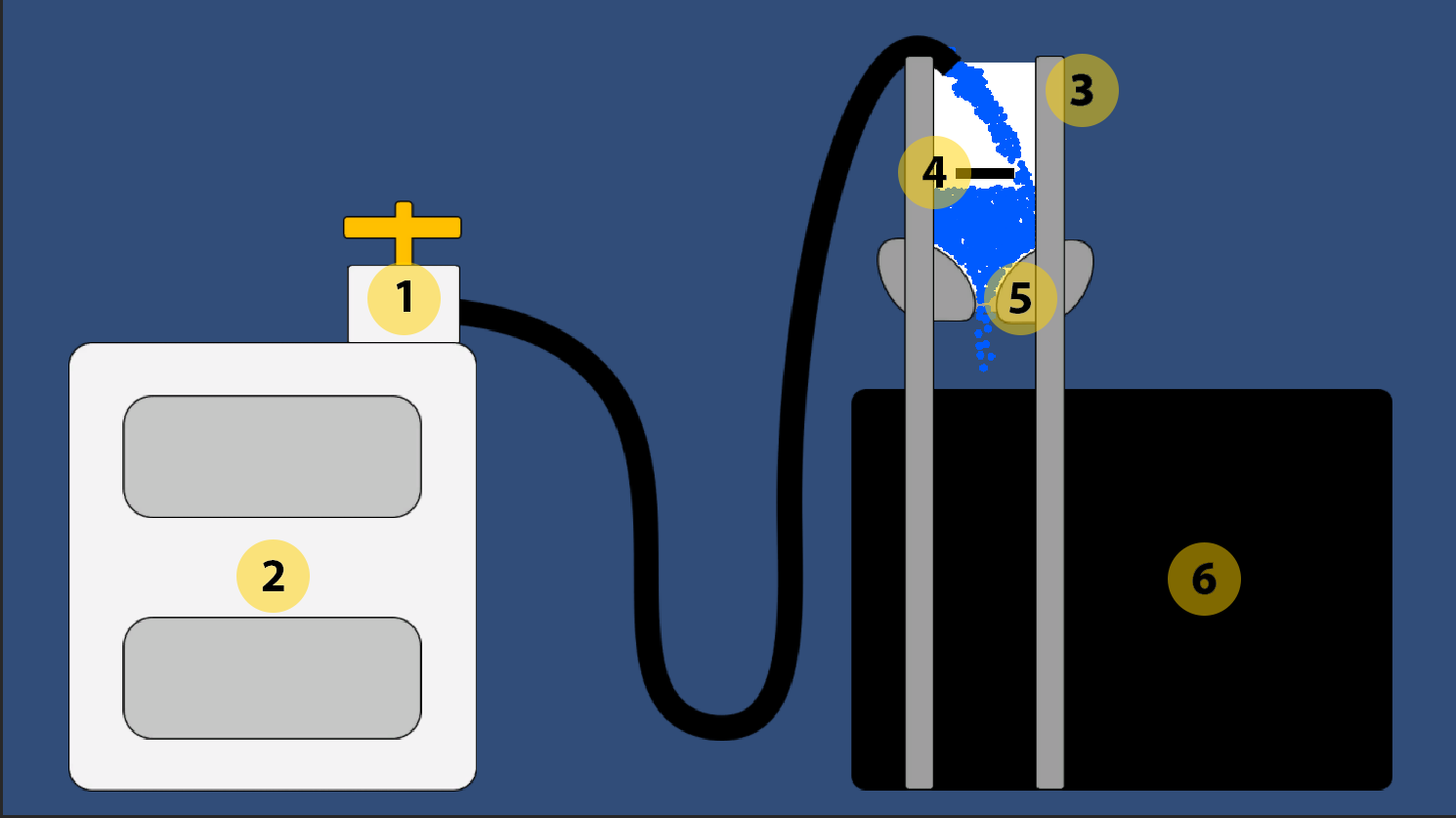}
    \subcaption{Virtual. All parts of the physical setup are simulated in a corresponding 2D environment.}
    \label{fig:setup_vis}
    \end{subfigure}
    \caption{Comparison of the physical and virtual representation. The pump (1) on the canister (2) represents the basic heart function. Liquid is transported to the buffer vessel (3) on the right. The level indicator (4) marks the point at which the water level has to be kept to represent 5--6 L/min of flow. The valve (5) leads into the reservoir (6).}
    \label{fig:setup}
\end{figure}

\noindent\textbf{Representation Design} We now proceed to design our models around the pump metaphor for the human heart. 
As mentioned above, the basic heart function is the transport of blood through the circulatory system. Physically, we choose to represent the human heart with a commercially available pump, suited to transport the same amount of liquid. The physical pump is operated by moving a piston up and down, which first creates negative pressure in the inlet when it is moved upwards. This is similar to the \textit{diastole} of the cardiac cycle. 
When the piston is moved downwards, the pressure inside the pump, towards the outlet is increased. This resembles the \textit{systole}.
The demand for blood in the body is represented by a buffer that is drained at a constant rate. This rate resembles the pump output of the heart. To keep the buffer filled, constant operation of the pump is required. In this state, the system represents the normal function of a heart. This manually operated, physical representation (\textbf{\PM}) is shown in Figure~\ref{fig:setup_phys}.

To compare manual and automated representations, we ensure that operating the pump can be taken over by a set of sensors and actors. While it would be preferable to leave the manual pump in place and design a system for automatic operation, we opted for a simpler design in favor of practicality.
We utilize an electric pump operating similarly to the manual one. Electrical pumps are typically designed for continuous operation as opposed to our chosen manual exemplary. However, this would destroy the systole/diastole metaphor. Therefore, we split the pump into one simulated and one operational part. The simulated part consists of a motor, operating a piston complete with simulated handles in reciprocating motion, to imitate the operation of a user. The operational part consists of the electrical pump, switched on and off in the rhythm of the simulated pumping. The system is controlled by a level sensor in the buffer vessel, which switches on the pump when the water level in the buffer vessel is under a certain threshold. Thus, we obtain a physical representation that works autonomously \textbf{(\PA)}.

After designing the physical representation, we now reproduce the pump metaphor in a virtual setting. To keep the interaction with the model similar to the physical installation, we simulate the process using Unity \cite{noauthor_unity_nodate}, since it provides a basic physics simulation environment that serves our purposes.
For the simulation, we take multiple considerations into account. 
It could be done in a 3D environment, where interaction is still performed via mouse and keyboard while introducing additional complexity for observers. 
Virtual and augmented reality simulations of the process could be used as well. These technologies present additional complications that could influence our metrics.
We opt for a 2D representation to limit interaction with the virtual representation to the same interaction points that exist in the physical version. In both representations, the handle of the pump is moved up and down to collect and expel liquid, and the targeted flow rate is symbolized by the watermark in the draining vessel. 
This way we limit the focus of our study to the factors of manifestation and mode of operation.

In our Unity simulation, we show a schematic view of the physical installation. The canister, pump, and buffer are represented by 2D objects, as shown in Figure~\ref{fig:setup_vis}. The piston is in the same position, on top of the canister, operated by moving the piston up and down. An observer can move the piston with mouse and keyboard gestures, while the drag and weight of the object are simulated in the physics engine. When the piston is moved downwards, the ejected water is represented by particles exiting the hose into the buffer. The particles exit the buffer from the bottom, depending on how wide an adjusted opening is, simulating the draining of the buffer. This represents the process virtually, with a manual operation \textbf{(\VM)}. These are the same interaction points as in the physical installation. The motion of the piston can be automated using a simple script to obtain an automated virtual representation \textbf{(\VA)}, as the virtual counterpart of the automated pump. 

\begin{figure*}[ht!]
    
    \centering
    \begin{subfigure}[t]{0.24\textwidth}
    \centering
    \includegraphics[width=\textwidth]{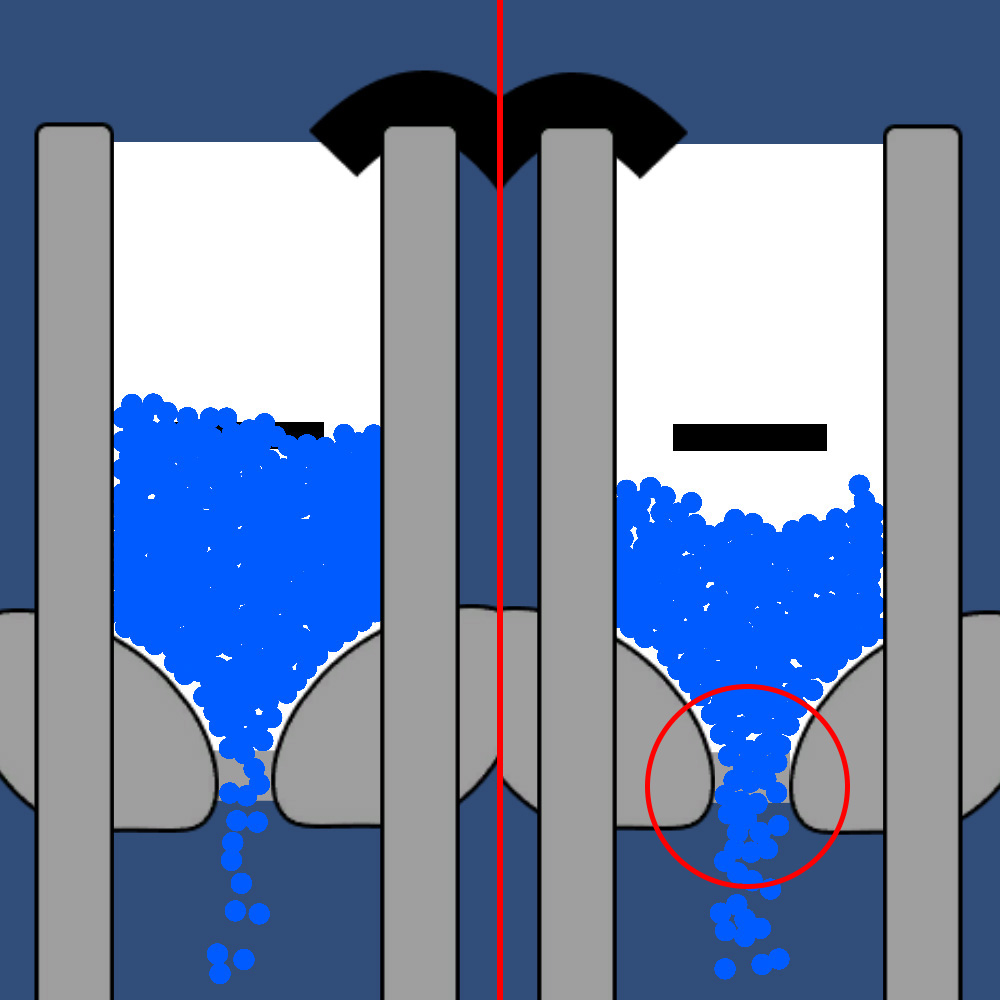}
    \includegraphics[width=\textwidth]{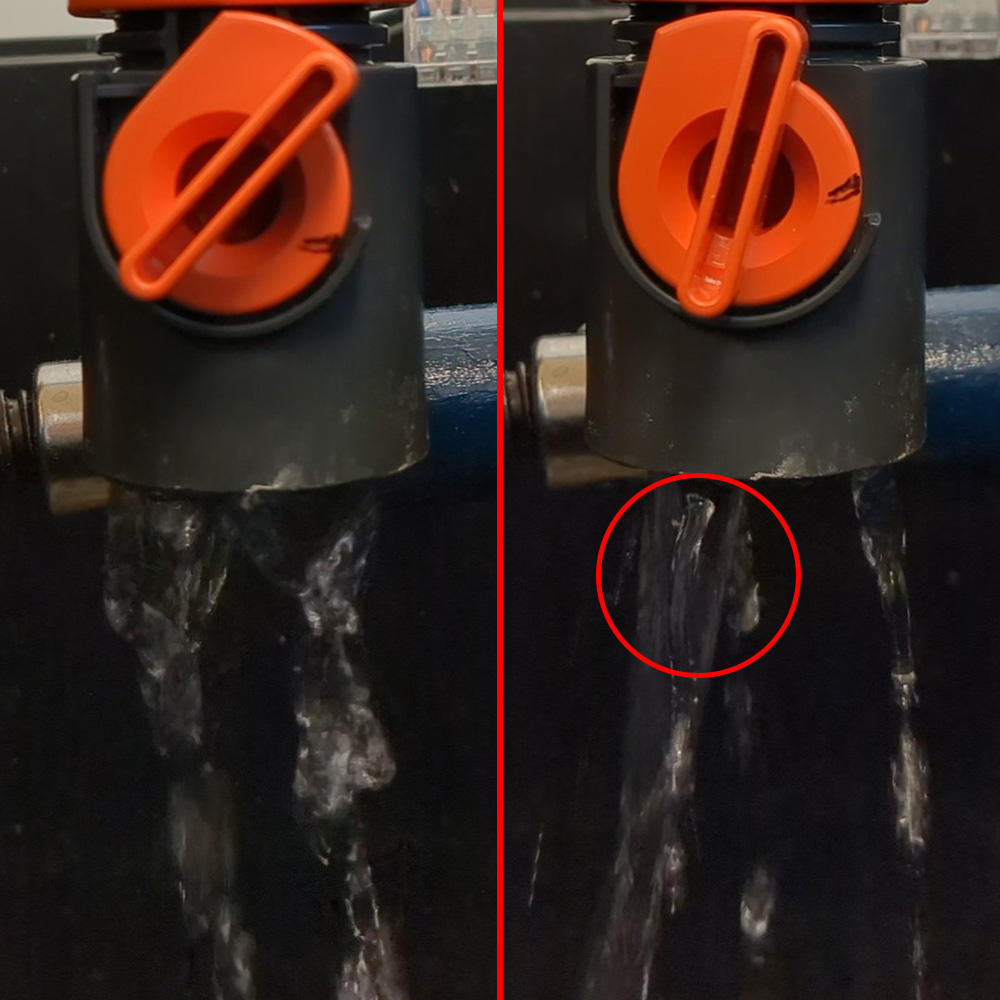}
    \subcaption{Exertion. Higher oxygen demand is represented by opening the drainage valve of the buffer vessel.}
    \label{fig:cond_exitation}
    \end{subfigure}
    \begin{subfigure}[t]{0.24\textwidth}
    \centering
    \includegraphics[width=\textwidth]{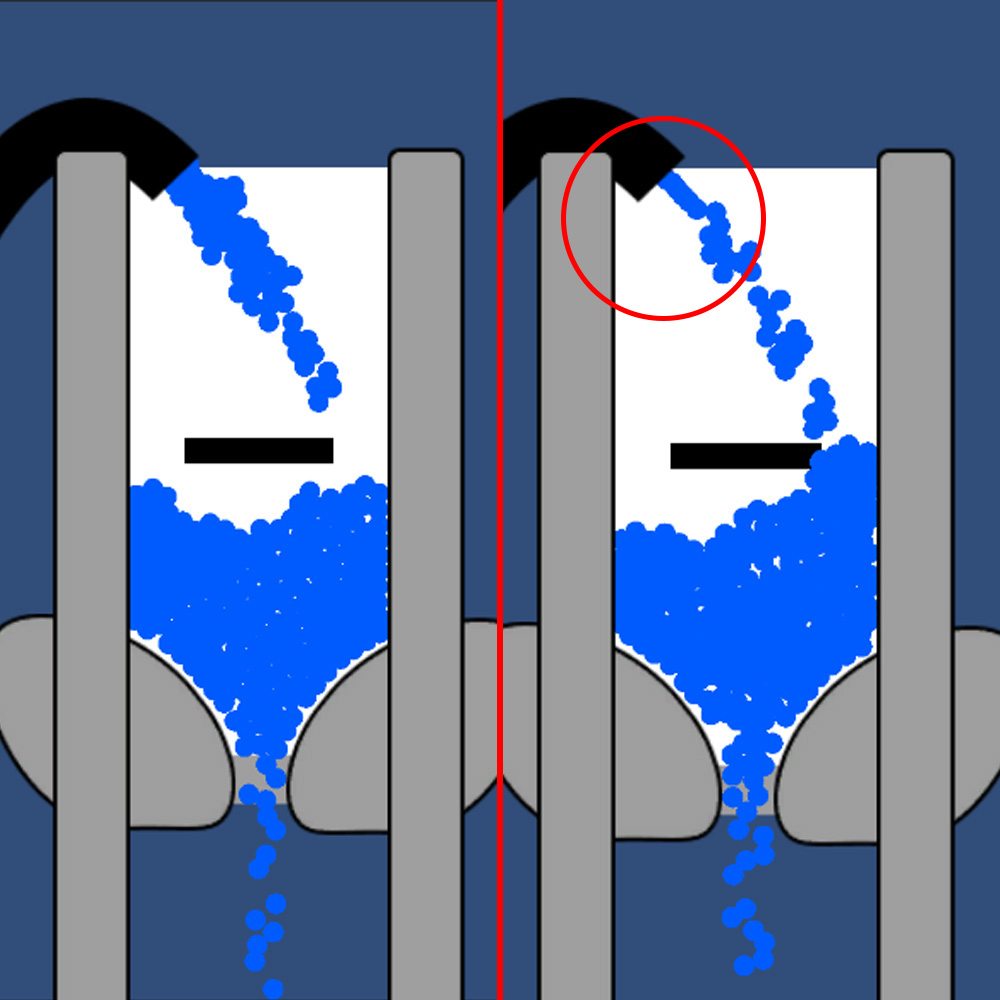}
    \includegraphics[width=\textwidth]{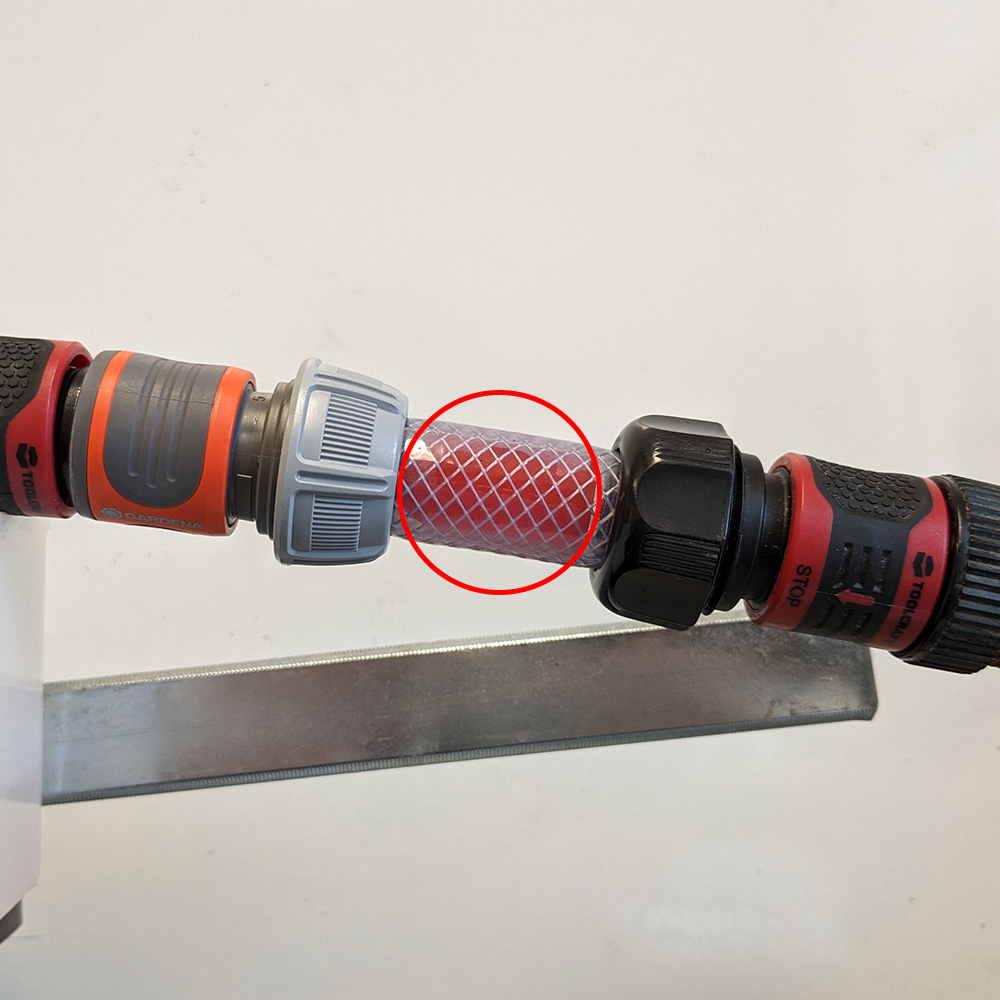}
    \subcaption{Aortic stenosis. The narrowing of the aortic valve is represented by a narrowing of the pump's outlet.}
    \label{fig:cond_aortic_stenosis}
    \end{subfigure}
    \begin{subfigure}[t]{0.24\textwidth}
    \centering
    \includegraphics[width=\textwidth]{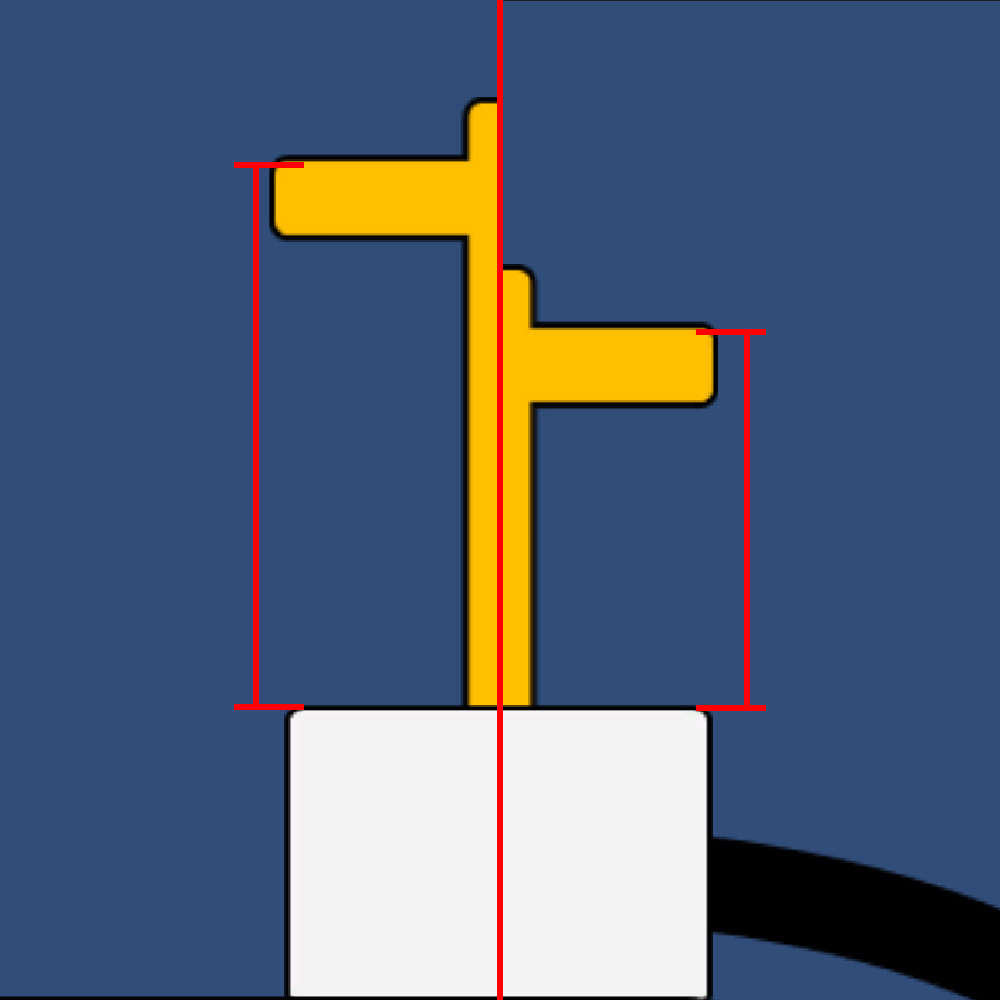}
    \includegraphics[width=\textwidth]{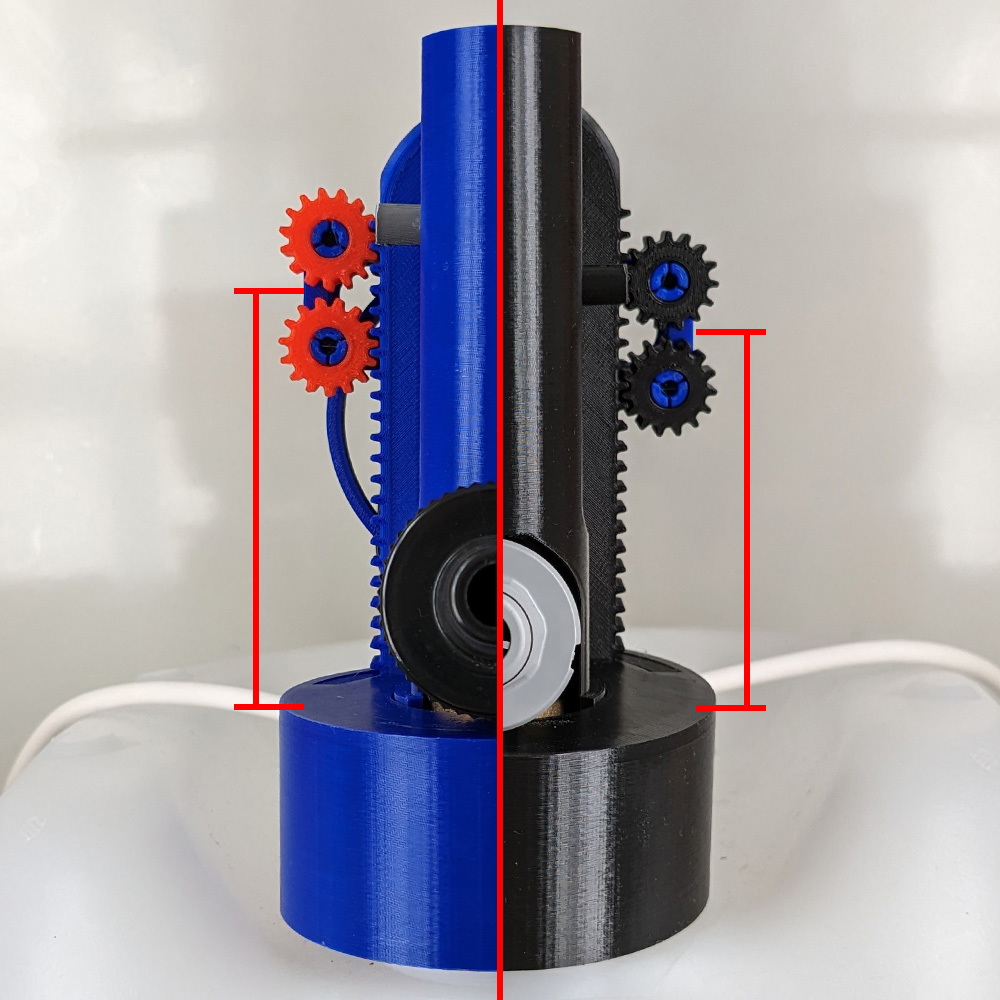}
    \subcaption{Systolic heart failure with reduced stroke volume. Limited heart volume is represented by limiting the movement range of the piston.}
    \label{fig:cond_systolic_fail}
    \end{subfigure}
    \begin{subfigure}[t]{0.24\textwidth}
    \centering
    \includegraphics[width=\textwidth]{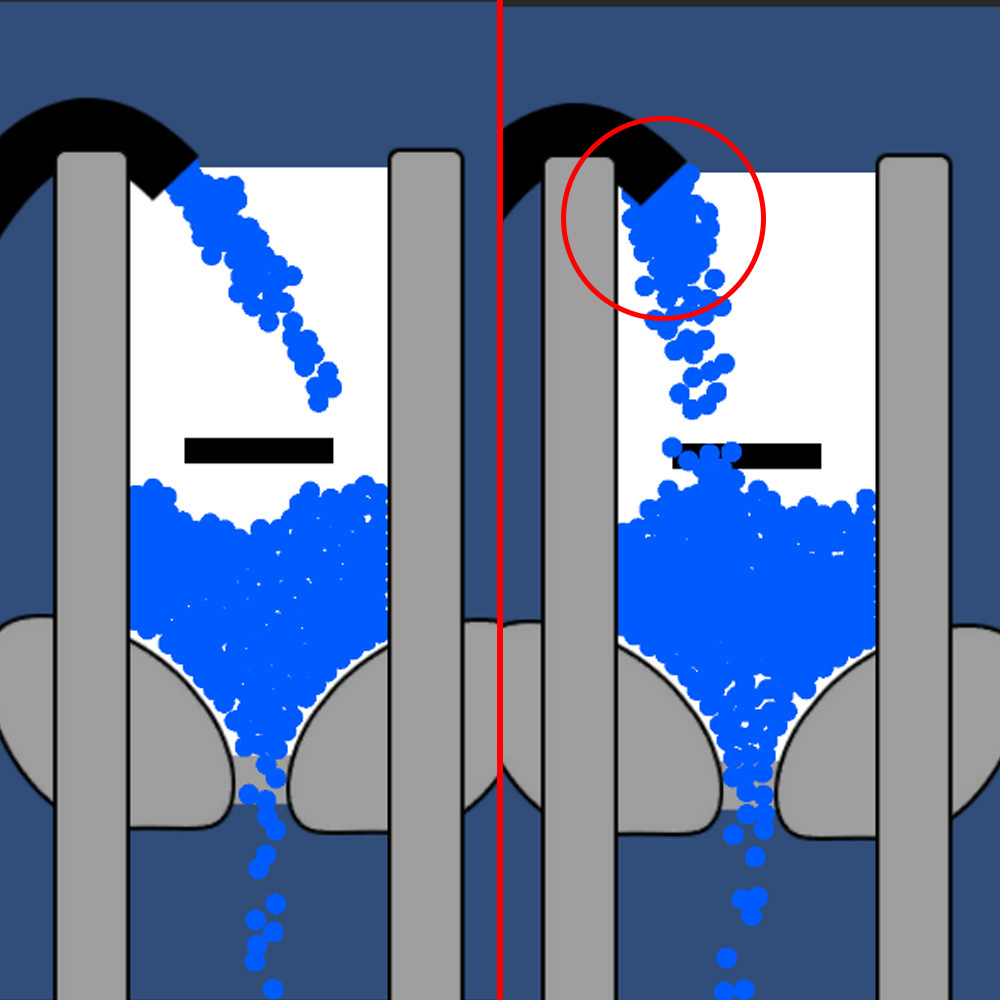}
    \includegraphics[width=\textwidth]{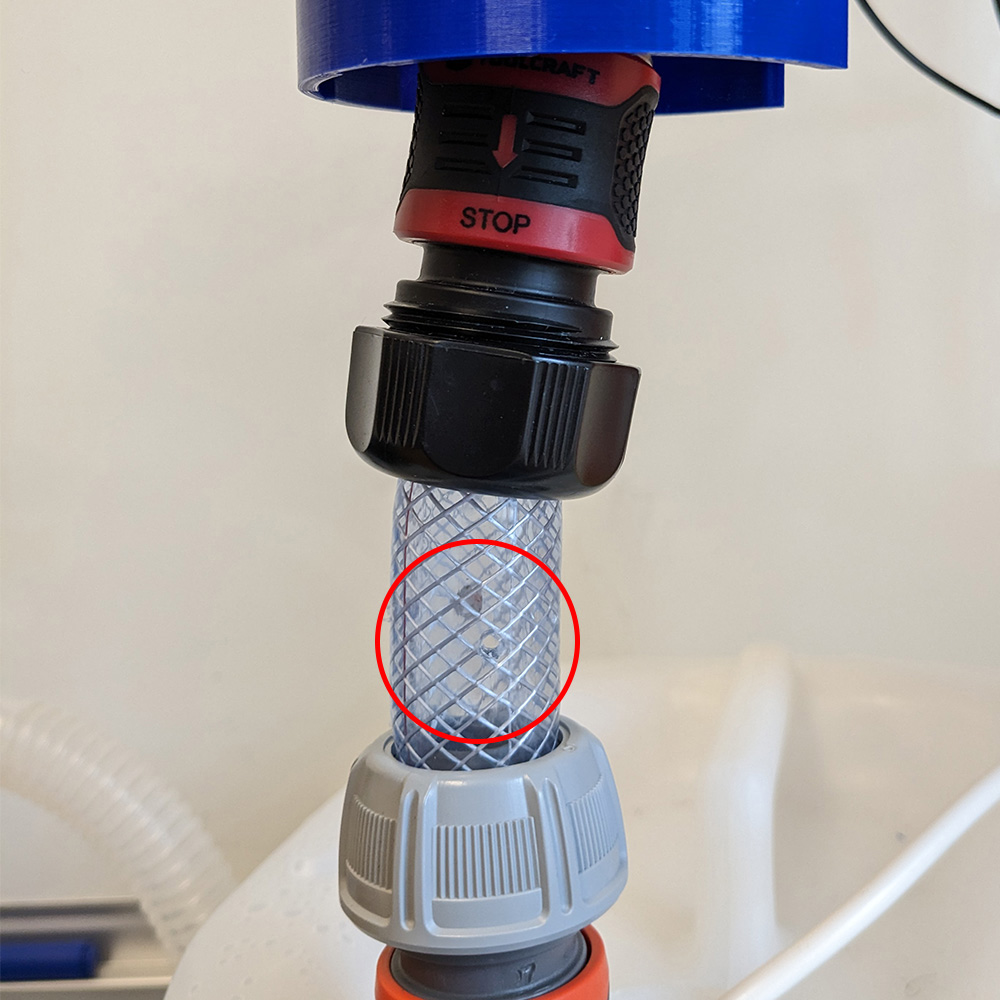}
    \subcaption{Cor Pulmonale. Difficulty in transporting blood to the lungs is simulated by puncturing the inlet of the pump.}
    \label{fig:cond_right_insufficiency}
    \end{subfigure}
    
    \caption{Scenarios, representing different heart conditions with a virtual (top) or physical metaphor (bottom).}
    \label{fig:conditions}
\end{figure*}

\subsection{Variations of the Model}
\label{ssec:variations}
For the representations of the cardiac function, as shown in Figure~\ref{fig:setup}, we now introduce different physiological and pathological states to create variations of the model. An overview of these scenarios and how they are represented in the physical and virtual representation respectively is shown in Figure~\ref{fig:conditions}. The individual scenarios illustrate how cardiac function changes in contrast to normal function. We use examples of common pathologies that can be illustrated using our model. In a consultation with a medical doctor, we ensured that the scenarios were depicted correctly, within the limited scope of our metaphor. 
In our study, participants are presented with each of the four variations of our representation (\PM, \PA, \VM, \VA), each depicting a different scenario. Their task is to determine the difference between normal cardiac function and the altered state in the specific scenario.
The four scenarios (\textbf{S1--4}), are discussed below: 

\noindent \textbf{Exertion (S1)} leads to increased demand for oxygen in the body. While at rest the cardiac output is $5-6$ $L/min$, this output can increase drastically during exercise, to even more than $35$ $L/min$ in elite athletes \cite{king_physiology_2022}. In our model, we represent the cardiac output by a slowly draining buffer vessel. In the interactive version, observers have to keep the buffer vessel filled around a marked threshold. In the automated version, a binary level sensor determines the fill state and pauses the pump when the sensor triggers. A state of exertion is simulated by opening the drainage valve of the buffer vessel further, as compared to the normal state. This is illustrated in Figure~\ref{fig:cond_exitation}.

\noindent \textbf{Aortic Stenosis (S2)} is caused by a stiffening of the aortic valve, leading it to not properly open. The heart has to work harder to contract the left ventricle and the systolic pressure increases \cite{pujari_aortic_2022}. To represent this physically, we 3D printed an obstruction of the pump's outlet, reducing the area of the outlet hose by half. In the virtual representation, this is simulated by increasing the resistance of the piston, 
as well as increasing the speed of the particles spilling into the reservoir. We illustrate the visible differences of this in Figure~\ref{fig:cond_aortic_stenosis}.

\noindent \textbf{Systolic Heart Failure (S3)} or heart failure with reduced ejection fraction is a condition where the left ventricle's pumping output is reduced. Here, the heart's ability to contract is limited by physical factors \cite{hajouli_heart_2022}. We represent the reduced pumping capability by limiting the movement range of the piston in the physical model. We automate the physical pumping process by simulating a reciprocating pump. The reduction of stroke volume is achieved by reducing the length of the rotating lever operated by the motor.  Virtually this is represented by limiting the range of the piston. This can be seen in Figure~\ref{fig:cond_systolic_fail}.

\noindent \textbf{Cor Pulmonale (S4)} is the fourth condition we simulate. In this case, the part of the heart that pumps blood into the pulmonary artery is affected. Depending on the cause, this does not entail a reduced ejection fraction \cite{voelkel_right_2006}. We represent pulmonary hypertension by drilling a hole into the inlet of the pumping mechanism. In the virtual representation, we simulate this by a reduction of the number of particles spawned by the outlet and a decreased weight of the piston. Figure~\ref{fig:cond_right_insufficiency} depicts this condition.

\section{Study}

We use a $2 \times 2$ full-factorial within-subject design, with manifestation and mode of operation as independent variables. Manifestation can be physical (\Ph) or virtual (\Vi), and operation can be manual (\Ma) or automatic (\Au). 
This results in four representations with different combinations of manifestation and mode of operation (\PM, \PA, \VM, \VA), which are illustrated in Figure~\ref{fig:teaser}.  
The scenarios were assigned to representations using a $4 \times 4$ Graeco-Latin Square design \cite{noauthor_graeco-latin_2008} to combat familiarization effects. In a sequence of four runs of the study, the order of scenarios, as well as the order of representations was unique. 
A number of participants (28) that is a multiple of four ensures that each combination of model and scenario is examined an equal amount of times.

\subsection{Participants}

We recruited $28$ participants for our study from faculty staff and students. $20$ identified as male, and eight as female. $16$ of them used visual aids. Age was collected in discrete intervals between $18$ and $54$ years. The largest age group was between $25$ and $34$ years old, which was $20$ participants in total.
Two participants self-reported to be well versed in anatomy, $14$ participants reported basic knowledge, $11$ reported themselves to be novices, and two participants reported no anatomical prior knowledge. One participant has a high school degree, $11$ have a bachelor's degree, $12$ have a master's degree, and four completed a Ph.D. level education.
All trials of our study took place in person, in the same environment. The participation was voluntary and no compensation was provided. Participants could opt out at any point if they felt discomfort.

\subsection{Procedure}
\label{sec:methodology:procedure}

Every participant interacted with one representation (\textbf{\PM, \PA, \VM, \VA}) per scenario (\textbf{S1--S4}), within a randomized order per individual subject. Each run of the trial started with the participants being led into the room where the experiments were conducted. 
The participants were handed a written introduction summarizing the nature of the tasks they were about to be presented with. We explained the procedure and clarified essential vocabulary.
Subsequently, the participants were asked to read an introduction to the first scenario and we explained the task.

For each scenario, participants were first presented with a representation of normal heart operation, to familiarize themselves with the model and how it represents normal heart function. They were allowed as much time as they desired with the normal functioning model. 
After that, we introduced the complication into the model, i.e., the variations discussed in Section~\ref{ssec:variations}.
The participants were allowed as much time as they wanted with the altered state.
Then they were asked to complete a quiz, to determine how much they understood a complication (\textbf{H1}).
After the quiz, we asked them to complete two additional questionnaires.

\noindent\textbf{The task} was to answer questions about how the different complications affect certain vital parameters in a multiple-choice quiz.
The participants were presented with one representation, first in a normal state, then in a complicated state.  
Then they had to decide how the vital parameters \textbf{heart rate}, \textbf{cardiac preload}, and \textbf{afterload} were affected in the scenarios. The multiple-choice quiz allowed the answers \textbf{increased}, \textbf{lowered}, and \textbf{unaffected}. The quiz also contained binary questions about the presence of the symptoms \textbf{fatigue} and \textbf{shortness of breath}. We allowed the participants to indicate they are \textbf{unsure} about the answers, to prevent them from making random choices.
These parameters were the same for every scenario and were introduced and thoroughly explained to subjects beforehand. 

\noindent\textbf{The first questionnaire} the participants were confronted with was the NASA Task Load Index (TLX) \cite{hart_development_1988}. 
We closely kept to the instructions provided by NASA with only one exception: We used a discrete scale from 0 to 10 for the questionnaire as opposed to the 0--100 scale with 21 increments that NASA suggests. 
The TLX uses $6$ components to measure the subject perceived load during a task. The individual components are mental demand, physical demand, temporal demand, performance, effort, and frustration. These individual factors are measured on a scale from high to low (except for performance which is measured from good to bad). We used the resulting task load index to test for \textbf{H2}.
During the first experiment of every trial, we performed the scale ranking procedure for the TLX. This procedure is used to identify participants' subjective biases for the different components of the task load index. The task load index for a given task was then calculated by computing the weighted average of the participant's ratings. 

\noindent\textbf{The second questionnaire} assesses emotional engagement, as proposed by Wang et al. \cite{wang_emotional_2019}. They propose to measure engagement as comprised of the following categories: creativity, affective engagement, physical engagement, intellectual engagement, and social engagement. 
The items of the questionnaire consist of statements the participants rate on a scale of 0 to 10, to be consistent with our implementation of the NASA-TLX. To keep the questionnaire reasonably concise we used a single question for each factor of engagement as proposed by Wang et al. Similar to the TLX, we computed the engagement score as the average of all components, however, we performed no scale-weighing procedure for our engagement score. This score was used to test for \textbf{H3}.

\noindent Additionally, after every experiment, we asked the participants to leave \textbf{written feedback}.
While the participant fills the questionnaires we prepared the next experiment. The process was repeated until the participants had seen all four representations and filled out the corresponding questionnaires.
Subsequently, the subjects were asked to fill out a final questionnaire about their overall experience. In it, they are asked to rank the models from most to least enjoyed. This ranking was used in addition to the engagement score to test for \textbf{H3}.
The final questionnaire also contained two \textbf{open text questions} about the most and least like experiences with any representation overall. The written statements were used for the qualitative analysis.

Subjects were provided instructions for both parts of the questionnaire, with written explanations of every item, which they could refer to at any point. 
At the end of the session, there was a debrief, where we asked the participants about their overall thoughts on their experience with the models, and thanked them for their participation. 

\subsection{Analysis}
We used a mixed methods approach to address our three hypotheses. This section describes the analysis performed on both quantitative and qualitative data.

\noindent\textbf{Quantitative Analysis} We collected the results of the quiz (\textbf{H1}), NASA TLX (\textbf{H2}), and our custom engagement questionnaire (\textbf{H3}) for all experiments of all study participants. To analyze the influence of these two factors on the variables quiz-score, TLX-score, and embodiment score, we opt to employ Wobbrock et al.'s Aligned Rank Transformed ANOVA (ART-ANOVA) \cite{wobbrock_aligned_2011}; a non-parametric framework reliant on an initial rank transformation of the data that allows for the subsequent use of complete two-way ANOVA models. 
This non-parametric model was used because the model's assumption of normally distributed residuals was rejected when probed with Shapiro-Wilk tests.
The statistical significance of the factors was evaluated for each such model using ANOVA-standard $t$-tests.
Multiplicity was accounted for by adjusting the significance threshold using a standard Bonferroni correction.
For \textbf{H3} we additionally investigate participant's preference in a model's manifestation and mode of operation.
Participants ranked their personal preferences for each of the four combinations of operation mode and manifestation. 
These ranks, owing to their obvious non-normal distribution, were tackled non-parametrically.
More specifically, an initial omnibus Friedman test \cite{friedman_use_1937} was performed to investigate whether any of the four combinations were statistically significantly different from each other, in which case it was followed by a series of Wilcoxon Signed-Rank Tests~\cite{woolson_wilcoxon_2008} to probe their pairwise differences.
For these pairwise tests, multiplicity was again accounted for using a Bonferroni correction of the significance threshold across the six comparisons performed.

\noindent\textbf{Qualitative Analysis}
We used a coding approach to quantify our collected qualitative data \cite{chi_quantifying_1997}, with \textbf{three independent coders}.
For this, we broke down the collected written feedback in our questionnaires into $78$ utterances. In our qualitative analysis, we performed both deductive and inductive coding. 
As a first step, we assigned each comment to the representations it referred to. This was decided based on the context given by the questionnaires and the content of the statement. Multiple assigned representations were possible, for example, if a statement read ``\emph{\dots in the physical variants\dots}'', we assigned it to both of the physical representations.
We found some statements referring exclusively to our used metaphor, which would apply to every representation. These will not be used to validate the hypotheses and will be discussed separately. 
In the \textbf{deductive coding} step, we had three individual coders perform an assignment of the statements to the aspects of understanding (\textbf{H1}), task load (\textbf{H2}), and subjective enjoyment (\textbf{H3}) while rating them as positive or negative. Monitoring the Krippendorf $\alpha$ value \cite{krippendorff_computing_2011} for inter-coder agreement, we iteratively refined the assignment until an $\alpha$ greater than $0.8$ was reached.
The disputed cases were then decided by a majority vote.
Finally, each coder performed an \textbf{inductive coding}, assigning common concepts to their choices. These codes were unified in a final session.
We use this categorization to draw further insights from the feedback.
\section{Results} 
\label{sec:result}

In this section, we present the findings of both the quantitative and qualitative parts of our study. We then draw our conclusions in relation to each of our hypotheses in a triangulation.

\subsection{Quantitative Results}

\noindent\textbf{Understanding (H1)}
The quiz scores yielded diverse results across both manifestations and interaction modes.
The ART-ANOVA conducted showed \textbf{neither a main effect for manifestation} ($F(1,27)=0.025; p=0.87$) \textbf{nor for operation} ($F(1,27)=1.28;p=0.27$). 
We could also not show an interaction effect between operation and manifestation to be statistically significant ($F(1,27)=0.01;p=0.92$).

\begin{figure}[t]

    \centering
    \includegraphics[width = 0.48\textwidth]{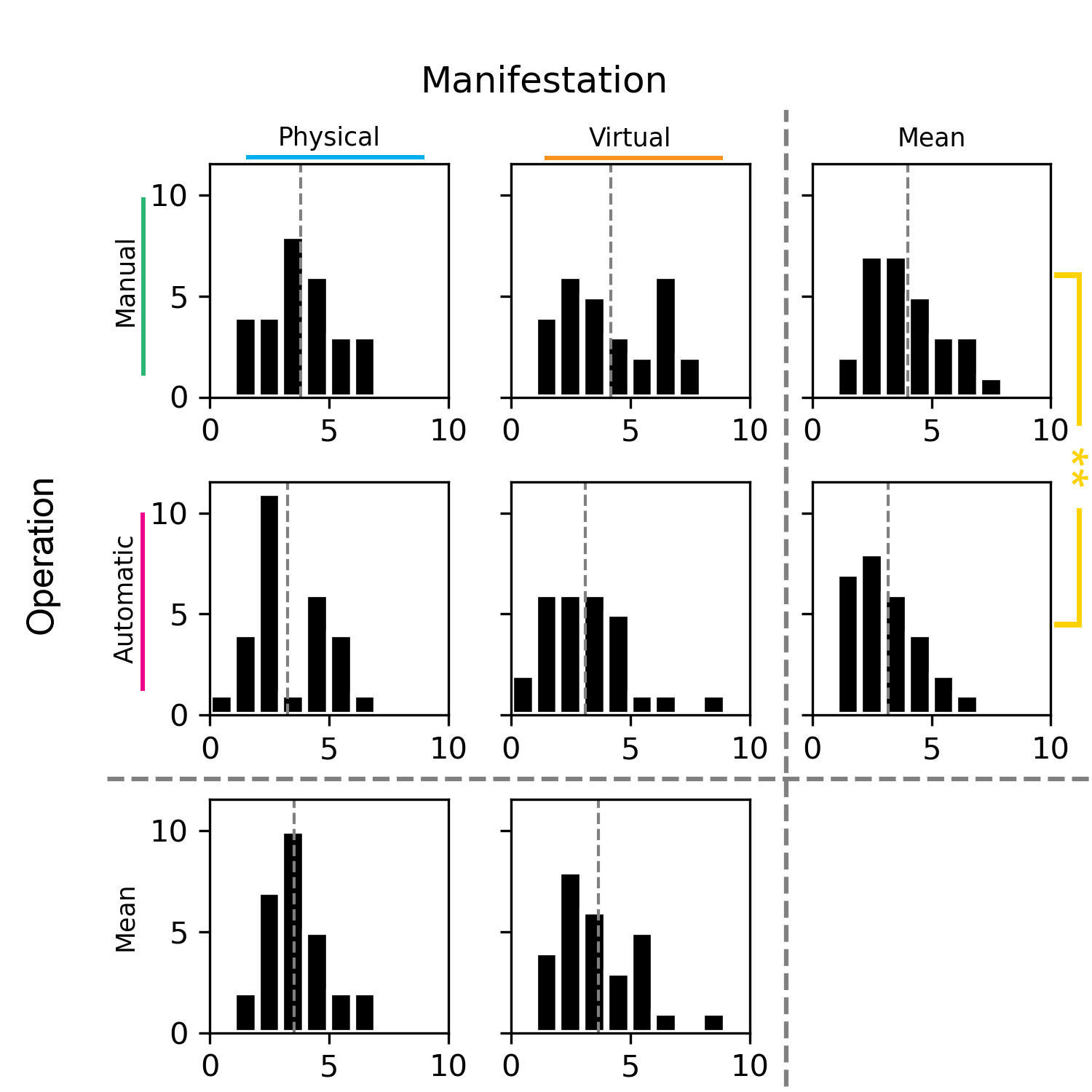}    
    \caption{Histogram matrix of TLX scores by manifestation and operation mode.  Average scores per participant across dimensions are shown behind the dashed lines. The \mc{manual} row shows an increased task load.}
    \label{fig:plt_tlx}
    
\end{figure}

\noindent\textbf{Task Load (H2)}
The weighted NASA-TLX scores are shown in Figure~\ref{fig:plt_tlx}.
Firstly, the ANOVA showed no significant effect of manifestation on task load ($F(1,27)<0.01;p=0.99$). 
However, the ANOVA highlighted a \textbf{significant effect of \mc{manual} interaction on task load} ($F(1,27)<7.94;p<0.01$).
Participants, as expected, reported higher physical demand in the interaction with \pc{physical} representations (median=$3$) compared to \vc{virtual} ones (median=$1.5$).
When comparing task load for different operation modes, \mc{manual} models have a notably higher physical demand component (physical demand median=$4.5$) than \ac{automated} ones (physical demand median=$0$). 
The effort component was also consistently rated higher for \mc{manual} models (median=$5$) than for \ac{automated} ones (median=$3.25$). 

\begin{figure}[t]

    \centering
    \includegraphics[width = 0.48\textwidth]{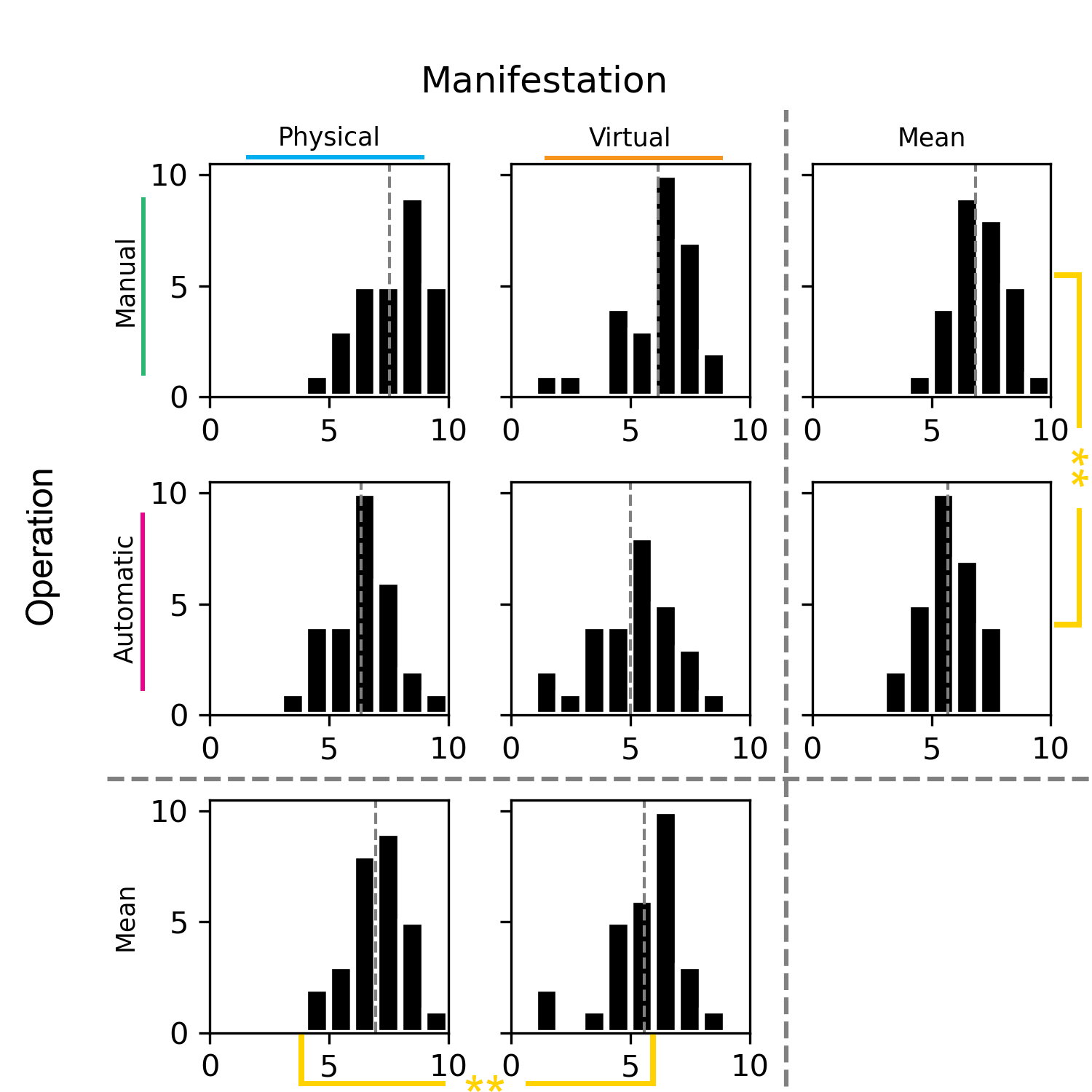}    
    \caption{Histogram matrix of engagement scores by manifestation and operation mode. Average scores per participant across dimensions are shown behind the dashed lines. Both \mc{manual} and \pc{physical} sides show an increase in engagement.}
    \label{fig:plt_engagement}
    
\end{figure}

\noindent\textbf{Emotional Engagement and Preference (H3)}
Figure~\ref{fig:plt_engagement} shows the results of the engagement questionnaire.
Here, the conducted ANOVA showed a \textbf{statistically significant effect of manifestation} on the engagement score ($F(1,27)<18.73;p<0.01$). The median engagement score for \pc{physical} models is $7$, while for \vc{virtual} ones it is $5.95$.
The \textbf{effect of \mc{manual} interaction on engagement was also statistically significant} ($F(1,27)<31.61;p<0.01$), with the overall score for \mc{manually} operated representations ($6.95$) higher than that of \ac{automatic} ones ($5.75$).

When comparing engagement scores of \pc{physical} and \vc{virtual} representations in detail, \pc{physical} models have produced higher scores across several engagement components.
More specifically, creativity, affective engagement (medians $6.5$ vs $5$), social engagement (medians $7.5$ vs $6.75$), as well as physical engagement (medians $6$ vs $4.25$) all receive higher median ratings for \pc{physical} representations. 
Comparing engagement scores between the two modes of interaction, \mc{manual} representations receive a noticeably higher physical (medians $7.5$ vs $4$), social (medians $7.75$ vs $6$), and affective (medians $6.25$ vs $5$) engagement score.
We also did not find a significant correlation between task load and engagement.


\begin{figure}[b]
    \centering
    \includegraphics[width = 0.48\textwidth]{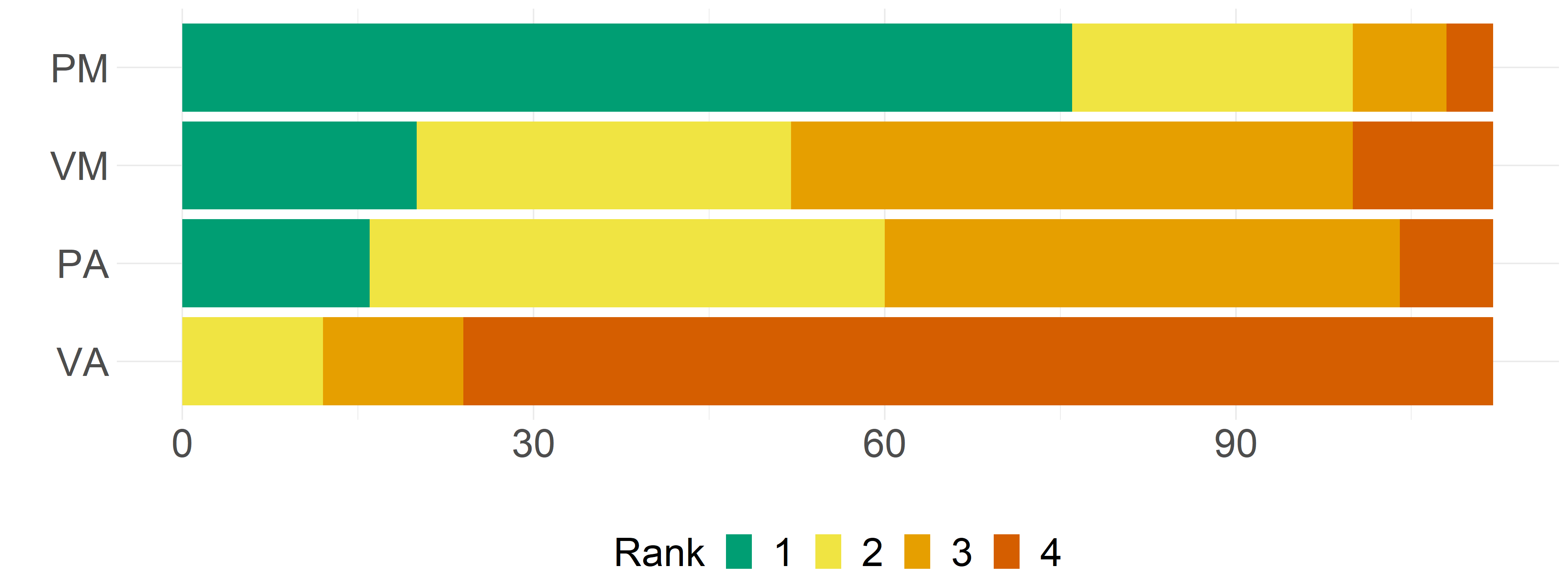}
    \caption{Stacked bar charts of preference votes for the different models. All participants chose their most (green) to least preferred (red) representation in the final questionnaire. 
    \pc{Physical} \mc{manual} representations are significantly more preferred than others, while \vc{virtual} \ac{automated} are the least popular.}
    \label{fig:plt_preference}
\end{figure}
We finally show the results of the ranking of the four representations in Figure~\ref{fig:plt_preference}. 
To probe this hypothesis quantitatively, we performed an omnibus Friedman Rank Sum Test to compare the distribution of ranks across the four combinations of manifestation and operation mode ($\chi^2= 41.571$, $p<0.01$). 
Given the Friedman test's significance, we subsequently performed paired, post-hoc Wilcoxon Rank Signed Tests to investigate each of the six pairwise differences between models.
We observe significant differences between most variants, with the notable exception of \pc{physical} \ac{automated} and \vc{virtual} \mc{manual} representations ($p=1$).

\subsection{Qualitative Results}

\begin{figure}[h]
    \centering
    \includegraphics[width = 0.49\textwidth]{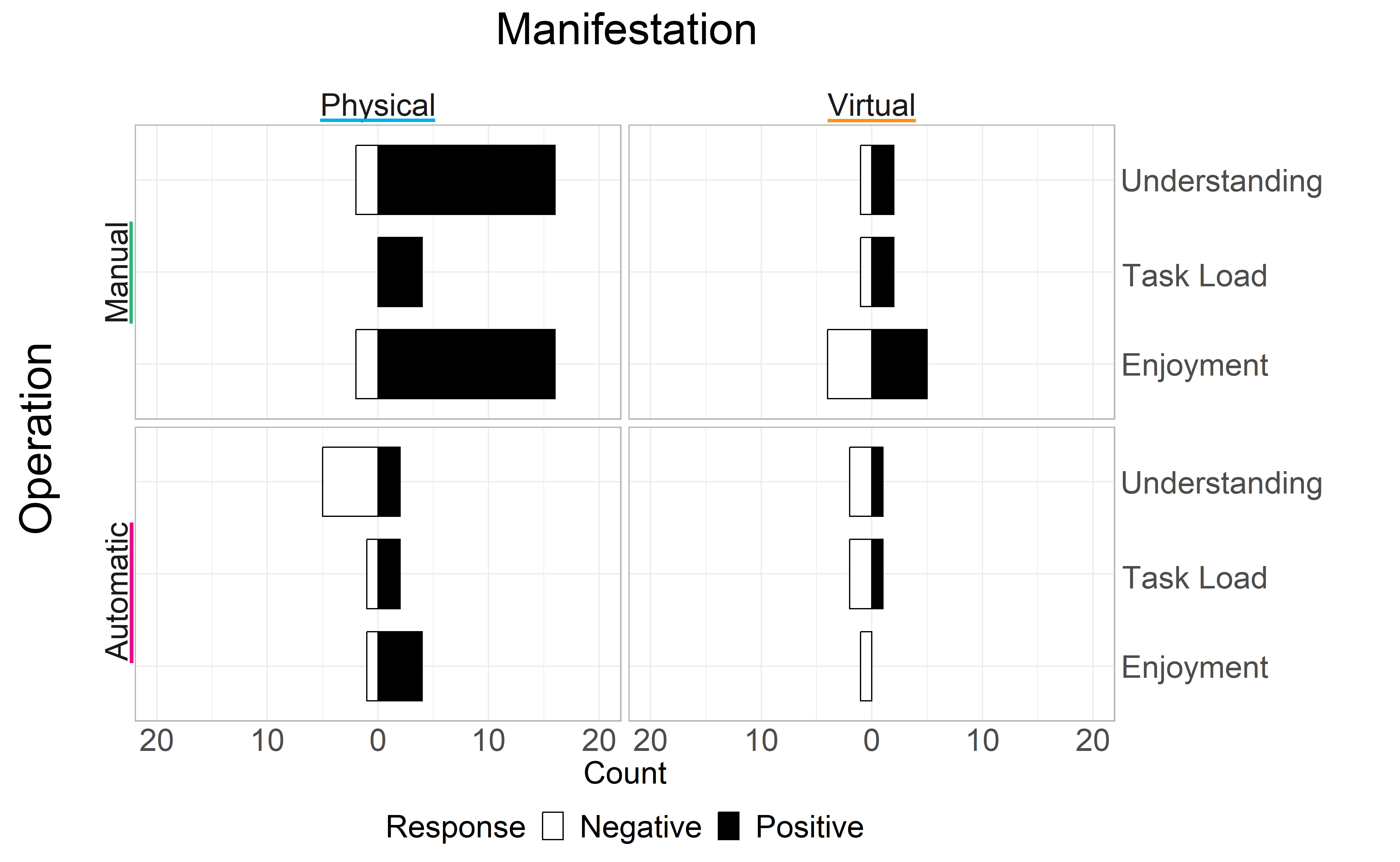}
    \caption{Results for the deductive coding of the qualitative feedback. The Y-axes in the subplots refer to the codes of the respective hypotheses.}
    \label{fig:plt_coding}
\end{figure}

We present a summary of the result of the deductive coding step in Figure~\ref{fig:plt_coding}. 
The codes that emerged in the inductive coding are explained below, in conjunction with the associated deductive codes. Selected examples for statements are given in quotes.

\noindent\textbf{Understanding (H1)} 
We found $16$ statements that described increased understanding when using \pc{physical} \mc{manual} representation, but only two negative statements on this. We registered six statements reporting \textbf{non-visual insight}, i.e. insights not gained over the visual channel, five of which directly referenced added haptic information (``\emph{[i liked the] haptic feedback of the pressure on the pump}''), and one referring to audible changes in the system. Additionally, six statements reported \textbf{visual insight} gained from \pc{physical} \mc{manual} representations (``\emph{seeing the flow rate as water gave me a good idea of how the volume moved}''). 
Conversely, the \ac{automated} representations with \pc{physical} \mc{manifestation} received five and \vc{virtually} manifested two negative comments respectively. Here, participants reported confusion about how \ac{automated} representations would react to complications (``\emph{frequency was missing}", ``\emph{visualization of the diastole was unclear}'').

\noindent\textbf{Task Load (H2)}
Deductive coding reveals four statements pointing towards increased task load when working with \pc{physical} \mc{manual} representations. The inductive coding shows three of these comments referring to increased \textbf{cognitive effort}. Examining these comments in detail, we find that only one participant specifically stated that they found added cognitive load added by the \pc{physical} \mc{manual} representation (``\emph{The hands-on interaction adds an extra layer of cognitive effort}''). One participant reported difficulties in determining the difference between \pc{physical} \mc{manual} and \ac{automated} representation. The final comment was about the thickness of the black line, and finding it hard to see the water line. For the \vc{virtual} \mc{manual} representation, two participants stated the desire for refined controls. The only statement about \textbf{physical effort} in the \pc{physical} representation states the necessity to step back to gain a better overview. 

\noindent\textbf{Enjoyment (H3)}
$16$ Participants made positive remarks on their enjoyment of the \pc{physical} \mc{manual} (`\emph{`Doing the work yourself was engaging}'', ``\emph{Manual pump was fun}'', ``\emph{[I liked the] interactivity}''), five for the \vc{virtual} \mc{manual}, and four for the \pc{physical} \ac{automated} representation. We did not receive any positive comments on the \vc{virtual} \ac{automated} model. Statements about \textbf{interaction} were a large group of positive remarks on both \mc{manual} representations, counting $15$ for the \pc{physical} and five for the \vc{virtual}. As another positive factor for the enjoyment of both of the \pc{physical} models, four participants reported positively on our design, which we grouped into the category \textbf{aesthetics} (``\emph{The clear tubes in the physical display were most compelling}'').

\subsection{Triangulation}
\arrayrulecolor{gray}

\begin{table}[t]
\caption{Summary of the triangulation process.}
\begin{tabular}{c|c|l|c}
                                    & \cellcolor[HTML]{C0C0C0}\textbf{Quantitative} & \multicolumn{1}{c|}{\cellcolor[HTML]{C0C0C0}\textbf{Qualitative}} & \cellcolor[HTML]{C0C0C0}\textbf{Triangulation} \\ \hline
\cellcolor[HTML]{EFEFEF}            &                                                                & \emph{Deductive Code:}                      &                                                                 \\
\cellcolor[HTML]{EFEFEF}            &                                                                & $\bullet$ Understanding                                                            & Qualitative                                                     \\
\cellcolor[HTML]{EFEFEF}\textbf{H1} & Quiz                                                           & \emph{Inductive Codes:}                      & support                                                         \\
\cellcolor[HTML]{EFEFEF}            &                                                                & $\bullet$ Non-visual understanding                                                 &                                                                 \\
\cellcolor[HTML]{EFEFEF}            &                                                                & $\bullet$ Visual understanding                                                     &                                                                 \\ \hline
\cellcolor[HTML]{EFEFEF}            &                                                                & \emph{Deductive Code:}                       &                                                                 \\
\cellcolor[HTML]{EFEFEF}            & NASA-                                                          & $\bullet$ Task load                                                                & Quantitative                                                    \\
\cellcolor[HTML]{EFEFEF}\textbf{H2} & TLX                                                            & \emph{Inductive Codes:}                    & support                                                         \\
\cellcolor[HTML]{EFEFEF}            &                                                                & $\bullet$ Physical load                                                            &                                                                 \\
\cellcolor[HTML]{EFEFEF}            &                                                                & $\bullet$ Cognitive load                                                           &                                                                 \\ \hline
\cellcolor[HTML]{EFEFEF}            &                                                                & \emph{Deductive Code:}                      &                                                                 \\
\cellcolor[HTML]{EFEFEF}            & Questionnaire,                                                 & $\bullet$ Enjoyment                                                                & Full                                                            \\
\cellcolor[HTML]{EFEFEF}\textbf{H3} & Ranking                                                        & \emph{Inductive Codes:}                    & support                                                         \\
\cellcolor[HTML]{EFEFEF}            &                                                                & $\bullet$ Interaction                                                              &                                                                 \\
\cellcolor[HTML]{EFEFEF}            &                                                                & $\bullet$ Aesthetics                                                               &                                                                
\end{tabular}
\label{tab:triangulation}\vspace{-10pt}
\end{table}

Finally, we discuss the triangulation of our results (Table~\ref{tab:triangulation}): 

\noindent\textbf{Understanding (H1)}
While quantitative methods did not show measurable improvements, the positive results for the inductive coding in the category \textbf{understanding}, as well as the positive results for \pc{physical} \mc{manual} representations in the inductive categories \textbf{visual understanding} and \textbf{non-visual understanding} positively support our theory.
In conclusion, H1 is only supported subjectively, and we can not show a measurable knowledge gain. Thus, we recommend further research on the effect of haptic feedback in \mc{manual} representations.

\noindent\textbf{Task Load (H2)}
Here, our quantitative analysis revealed that our participants reported a higher task load when working with \mc{manually} operated representations. However, our qualitative analysis does not fully support this result. Of the comments that relate to increased task load, only a single comment directly referred to a higher \textbf{cognitive effort} when working with \pc{physical} \mc{manual} representations in general. The remaining comments were specifically aimed at visual difficulties, resulting from the way we built our \pc{physical} model. The only cognitive difficulties in working with \vc{virtual} \mc{manual} models, reported only by two participants, referred to difficulties with the keyboard controls.  
\textbf{Physical effort} did not appear as a factor in the interaction with our models.
While there is some evidence to support the hypothesis that working with manual models increases subjective task load, we see little support for this from our qualitative analysis. We therefore conclude that task load did not play an important role for our participants. 

\noindent\textbf{Enjoyment (H3)}
In terms of the subjective enjoyment of \pc{physical} and \mc{manual} representations, the qualitative and quantitative results align. \pc{Physical} and \mc{manual} representations yielded significantly higher results on our engagement questionnaire. In the summary ranking, \pc{physical} \mc{manual} representations were preferred by a majority.
Further qualitative analysis reveals that \textbf{interaction} is a main facilitator of engagement, both in \pc{physical} and \vc{virtual} models. Additionally, the \textbf{aesthetics} of our representations were positively remarked upon in the \pc{physical} models.
The evidence collected supports the hypothesis that \pc{physical} manifestation, as well as \mc{manually} operated representations, are enjoyed more than \vc{virtual} or \ac{automated} ones.
\section{Discussion}
\label{sec:discuss}

Our study is a first attempt to disentangle the factors of operation mode and manifestation to highlight their individual effects on representations of a complex physiological process and associated complications. 
While we study a very specific example derived from the medical domain, we use common physicalization concepts like an embodied metaphor \cite{zhao_embodiment_2008} and direct interaction \cite{bae_making_2022}. With this, we aim to make the concept understandable to a lay audience.
From our results, we argue that \textbf{\pc{physical} manifestation and direct \mc{manual} operation are both beneficial factors in data physicalizations}.

Our quantitative metrics did not show a positive impact of \pc{physical} manifestation and \mc{manual} interaction on our participants' understanding of the represented process. This finding is consistent with prior studies conducted by Jansen et al.~\cite{jansen_evaluating_2013}, Hurtienne et al.~\cite{hurtienne_move_2020} and Drogemuller et al.~\cite{drogemuller_haptic_2021}.  However, our qualitative methods reveal a subjective benefit, as reported by participants in their feedback. This indicates that \textbf{\mc{manual} operation in the form of direct interactivity in \pc{physical} representations can lead to increased understanding}.
As opposed to \pc{physical} representations, \vc{virtual} representations do not benefit from such non-visual insight.
Interestingly, our findings contrast to the results by Pollalis et al.~\cite{pollalis_evaluating_2018}, who highlight the shortcomings of \pc{physical} representations in accurately representing visual information.
Instead, we find that when using \mc{manual} operation, \vc{virtual} and \pc{physical} representations can provide comparable levels of understanding. 
Finally, with respect to memorability, Stusak et al.~\cite{stusak_evaluating_2015} have shown that physicalizations are more memorable than virtual visualizations. Our study did not consider such long-term effects, but \textbf{we recommend investigating further the effects of direct interactivity on the memorability of physicalizations}.

Furthermore, we show that\textbf{ \pc{physical} data representations benefit highly from direct interactivity}. Our quantitative results indicate higher engagement for such representations, consistent with previous findings of Hurtienne et al.~\cite{hurtienne_move_2020}. Our results indicate that this is not caused by a difference in manifestation alone.
\textbf{If observers are not able to \mc{manually} interact with a representation, a \vc{virtual} one may be equally or better suited.} This is also supported by our participants' ranking of the different representations, where \vc{virtual} \mc{manual} and \pc{physical} \ac{automated} models were ranked similarly. 

In our study, we focus on metaphorical models. The creation of more complex representations would be more feasible in a purely \vc{virtual} setting than in the \pc{physical} space. \textbf{We think that \mc{manual} interaction can serve as an enhancement to both \pc{physical} and \vc{virtual} representations.}
Similar to Hurtienne et al.~\cite{hurtienne_move_2020}, who measured the physical effort to increase when working with a \pc{physical} representation, we observe increased reported task load in the \mc{manual} condition. On closer inspection, our quantitative results show that this may stem from the lessened use of cognitive resources of our participants when they did not interact with the representations themselves. Combined with the heightened engagement during \mc{manual} interaction, we conclude that \textbf{representations without direct interaction are less engaging.}

\section{Limitations and Future Work}

We designed our representation with the assumption that modeling the cardiac cycle as a simplified metaphor would be beneficial for a layperson audience.
Despite having been checked by a clinical doctor, our abstracted model and its variations may have caused misunderstandings of the (patho)physiological processes they represent.
In our study, we collected $14$ negative comments that referred to the model itself as opposed to aspects specific to manifestation or mode of operation.
Thus, \textbf{we recommend investigating further the effect of the degree of abstraction on (\pc{physical}) metaphors in future work.}

When creating our \vc{virtual} representations, we modeled the pump metaphor in unity with certain concessions. While using a simple, two-dimensional model of our \pc{physical} setup did not visibly impact our results, limiting the interaction with the virtual pump to the keyboard led to an absence of haptic feedback. 
Because of $15$ comments referencing the positive impact of \mc{manual} interaction on the engagement, as well as $6$ comments mentioning non-visual insights gained from \pc{physical} representations, a different method of controlling a virtual representation may yield more positive results.
Haptic feedback may be a greater influence than immersion, augmented or virtual reality methods may offer limited benefits.
\textbf{When comparing interactive representations in the future, we recommend using input devices with feedback mechanisms}. This could range from force-feedback joysticks to custom devices that simulate physical resistance.

Our analysis of the test scores failed to show an advantage of either of our four representations on the participant's understanding.
We used four scenarios to test our hypotheses with four different representations. One used scenario, simulating physical exertion (\textbf{S1}), was not fully equivalent to the other three.
Participants using manually operated representations scored worse in the exertion scenario.
This may have been because the possibility of non-visual insight was not given because of the lack of haptic feedback.
Therefore, \textbf{we recommend considering the presence of haptic feedback in \mc{manual} representations in future comparative studies.}

Finally, we acknowledge that our sample was biased towards a young, age group of male subjects with a relatively high education. Cardiac pathologies are relevant to everyone especially due to preventable risk factors, such as smoking. However, data physicalization may have additional benefits in educating an audience with low visualization literacy, including children. Future work considering a \textbf{more diverse demographic} could reveal further insights into the effectiveness of educational pathophysiological process representations---also with regard to learning aspects such as constructivism, active learning, or learning-by-doing~\cite{huron_constructing_2014, hogan_pedagogy_2017}.

\bibliographystyle{eg-alpha-doi}  
\bibliography{references}

\end{document}